\documentclass[9pt,letterpaper]{extarticle}  %,twocolumn
\usepackage{stmaryrd}
\usepackage{graphicx}
\usepackage{wrapfig}
\usepackage[latin1]{inputenc}
\usepackage{times}
\usepackage{color}
\usepackage{amsmath}
\usepackage{amssymb}
\usepackage{hyperref}
\usepackage{float}
\usepackage{dblfloatfix}
\usepackage{fixltx2e}

\usepackage{multicol}
\usepackage{caption}
\usepackage{setspace}

\usepackage[normalem]{ulem}

\usepackage{abstract}
\def \ua{{\uparrow}}
\def \da{{\downarrow}}
\def \be{\begin{equation}}
\def \ee{\end{equation}}
\def \ba{\begin{array}}
\def \ea{\end{array}}
\def \bea{\begin{eqnarray}}
\def \eea{\end{eqnarray}}

\def \half{{1\over 2}}

\def \a{{\alpha}}
\def \t{{\theta}}

\def \w{{\omega}}
\def \s{{\sigma}}

\def \yd{^\dagger}
\def \av#1{{\langle#1\rangle}}

\newcounter{firstbib}

\newenvironment{Figure}
  {\par\medskip\noindent\minipage{\linewidth}}
  {\endminipage\par\medskip}

\begin{document}

\begin{figure}[b!]
	\centering
	\includegraphics[width=150mm]{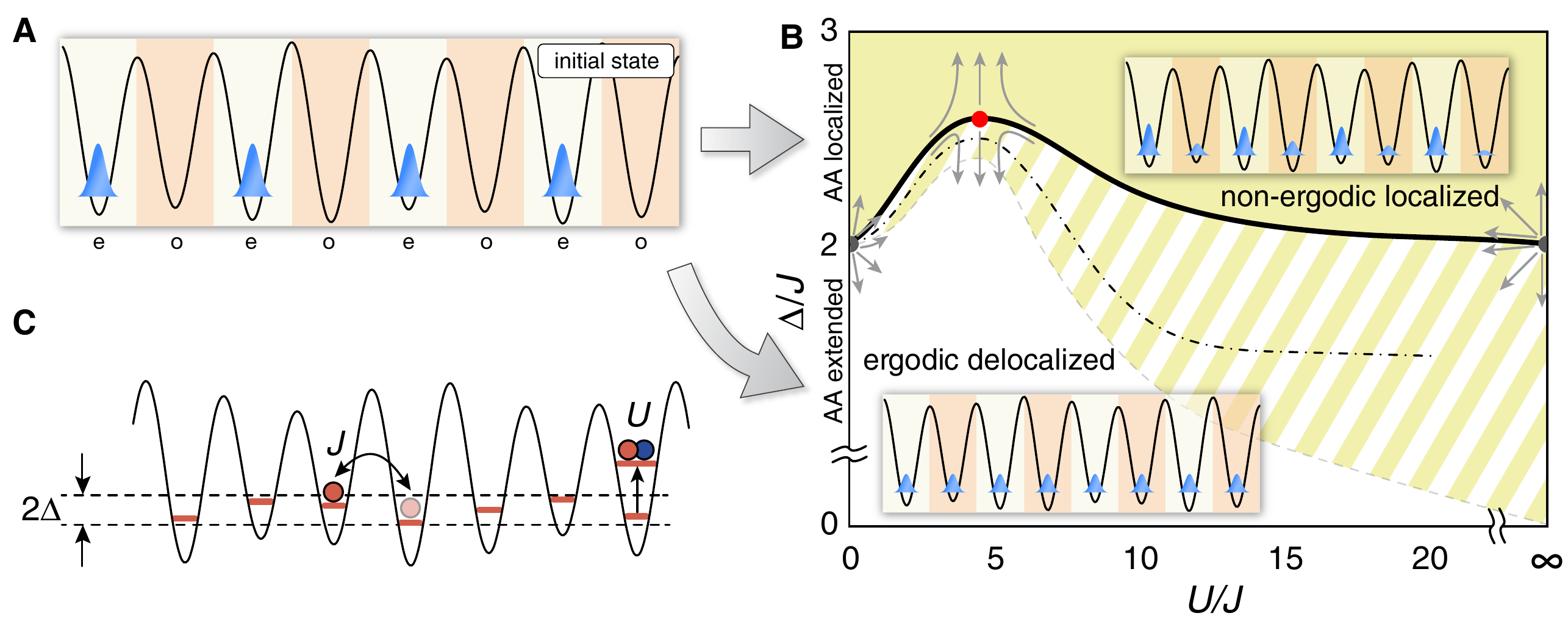} %76 %was 84mm
	\caption{{\small{\textbf{Schematics of the many-body system, initial state and phase-diagram.} \textbf{A.} Initial state of our system consisting of a charge density wave, where all atoms occupy even sites ($e$) only.  For an interacting many-body system, the evolution of this state over time depends on whether the system is ergodic or not.  \textbf{B.} Schematic phase diagram for the system: in the ergodic, delocalized phase (white) the initial CDW quickly decays, while it persists for long times in the non-ergodic, localized phase (yellow). The  striped area indicates the dependence of the transition on the doublon fraction, with the black solid line indicating the case of no doublons. The black dash-dotted line represents the  experimentally observed transition for a finite doublon fraction, extracted from the data in Fig.\ 4. The grey arrows depict the postulated pattern of renormalization group flows controlling the localization transition.  For $U=0$, as well as in the limit of infinite $U$ with no doublons present \cite{SOMs}, the transition is controlled by the non-interacting Aubry-Andr\'{e} critical point, represented by the unstable grey fixed points.  Generically, however, it is governed by the MBL critical point, shown in red.  \textbf{C.} Schematic showing a visual representation of the three terms in the Aubry-Andr\'{e} Hamiltonian (Eq.\ (\ref{AA_hamiltonian})).}}}
	\label{Schematic}
\end{figure}

\noindent
\textsf{\textbf{\Huge{Observation of many-body localization of interacting fermions in a quasi-random optical lattice
%Spins Take the Heat: Symmetry and localization protected topological q-bits at the edge of hot matter.
% title must be <90 characters
%Coherent manipulation of topological edge states at high temperature in a many-body localized  state\\
%OR: Coherent edge states protected by topology and localization from a  hot bulk\\
%OR: Quantum coherence at infinite temperature in a localized topological state
}}}

\vspace{0.5cm}

\noindent
\textsf{\large{\textbf{Michael\ Schreiber$^{1,2}$, Sean\ S.\ Hodgman$^{1,2}$, Pranjal Bordia$^{1,2}$, Henrik\ P.\ L\"uschen$^{1,2}$, Mark\ H.\ Fischer$^{3}$, Ronen\ Vosk$^{3}$,  Ehud\ Altman$^{3}$, Ulrich\ Schneider$^{1,2}$ and Immanuel\ Bloch$^{1,2}$}}}%LMU/MPQ/Weizmann$^{1,2,3}$ 

\noindent
\textsf{$^1$Fakult\"at f\"ur Physik, Ludwig-Maximilians-Universit\"at M\"unchen, Schellingstr. 4, 80799 Munich, Germany}

\noindent
\textsf{$^2$Max-Planck-Institut f\"ur Quantenoptik, Hans-Kopfermann-Str. 1, 85748 Garching, Germany}

\noindent
\textsf{$^{3}$Department of Condensed Matter Physics, Weizmann Institute of Science, Rehovot 76100, Israel}

%\begin{multicols}{3}
%
%\noindent
%(science report: text + abstract + captions + references up to about 2500 words, abstract not more than 125 words, about 30 references; 09.07.: text  + abstract  + captions  + references x = )

%\noindent
%(science article: text + abstract + captions + references up to about 4500 words, abstract not more than 125 words, about 40 references, up to 6 figs; 09.07.: text  + abstract  + captions  + references x = x)

\vspace{0.5cm}

\begin{abstract}

\noindent
We experimentally observe many-body localization of interacting fermions in a one-dimensional quasi-random optical lattice. We identify the many-body localization transition through the relaxation dynamics of an initially-prepared charge density wave. For sufficiently weak disorder the time evolution appears ergodic and thermalizing, erasing all remnants of the initial order. In contrast, above a critical disorder strength a significant portion of the initial ordering persists, thereby serving as an effective order parameter for localization. The stationary density wave order and the critical disorder value show a distinctive dependence on the interaction strength, in agreement with numerical simulations. We connect this dependence to the ubiquitous logarithmic growth of entanglement entropy characterizing the generic many-body localized phase.   

\end{abstract}

\begin{multicols}{1}

%\begin{multicols}{3}

\paragraph*{Introduction}

The ergodic hypothesis is one of the central principles of statistical physics. In ergodic time evolution of a quantum many-body system, local degrees of freedom become fully entangled with the rest of the system, leading to an effectively classical hydrodynamic evolution of the remaining slow observables \cite{Lux14}. Hence, ergodicity is responsible for the demise of observable quantum correlations in the dynamics of large many-body systems and forms the basis for the emergence of local thermodynamic equilibrium in isolated quantum systems \cite{Deutsch91, Sred94, Rigol08}. It is therefore of fundamental interest to investigate how ergodicity breaks down and search for alternative, genuinely quantum paradigms in the dynamics, and to understand the long-time stationary states that ensue in the absence of ergodicity. 

%Polkovnikov11, Lama12,    Peres84,     

One path to breaking ergodicity is provided by the study of integrable models, where thermalization is prevented due to the constraints imposed on the dynamics by an infinite set of conservation rules. Such models have been realized and studied in a number of experiments with ultracold atomic gases \cite{Paredes04,Kinoshita06,Gring12}. However, integrable models represent very special and fine-tuned situations, making it difficult to extract general underlying principles.

%,Ronzheimer13

Theoretical studies over the last decade point to many-body localization (MBL) in a disordered isolated quantum system as a more generic alternative to thermalization dynamics. In his original paper on single-particle localization, Anderson already speculated that interacting many-body systems subject to sufficiently strong disorder would also fail to thermalize \cite{Anderson58}. Only recently, however, have convincing theoretical arguments been put forward that Anderson localization remains stable under the addition of moderate interactions, even in highly excited many-body states \cite{Basko2006,Gornyi2005,Imbrie2014}. Further theoretical studies have established the many-body localized state as a distinct dynamical phase of matter that exhibits novel universal behavior \cite{Znidaric2008,Bardarson2012,Bauer2013,Vosk2013,Serbyn2013,Serbyn2013a,Huse2013a,Andraschko2014,Vasseur2014,Serbyn2014a,Lev14}. In particular, the relaxation of local observables does not follow the conventional paradigm of thermalization and is expected to show explicit breaking of ergodicity. In many ways, the MBL transition is fundamentally different from all other known transitions \cite{Vosk2014-2, Potter2015}.  On one side of the transition ergodicity prevails and quantum effects decay at long times, whereas on the other side quantum correlations persist indefinitely. Hence the MBL transition sets a sharp boundary between a macroscopic world showing quantum phenomena and one governed by classical physics.

%Vosk2014,Pekker2014,Bahri2013,Serbyn2014,

While Anderson localization of non-interacting particles has been experimentally observed in a variety of systems, including light scattering from semiconductor powders in 3D \cite{Wiersma97}, photonic lattices in 1D \cite{Lahini08} and 2D \cite{Schwartz07} and cold atoms in 1D and 3D random \cite{Billy08, Kondov11, Jendrzejewski12} and quasi-random \cite{Roati08} disorder, the interacting case has proven more elusive. Initial experiments with interacting systems have focused on the superfluid \cite{Deissler10,Derrico2014} or metal \cite{Kondov2013} to insulator transition in the ground state. Evidence for inhibited macroscopic mass transport  was reported even at elevated temperatures \cite{Kondov2013}, but is hard to distinguish from exponentially slow motion expected from conventional activated transport or effects stemming from the inhomogeneity of the cloud.  Until now conclusive experimental evidence for many-body localization at finite energy density has thus been lacking. 

%Initial experiments have studied global mass transport close to the ground state  in bosonic \cite{Deissler10,Derrico2014} and fermionic \cite{Kondov2013} ultracold atom systems  in the presence of disorder.  These localization effects are extremely difficult to observe in traditional condensed matter systems, as there will generically be some coupling to a phonon bath present which will destroy localization. While extremnly active theoretically, no experiments are available at high energy density .

In this paper we report the first experimental observation of ergodicity breaking due to many-body localization. Our experiments are performed in a one-dimensional system of ultracold fermions in a bi-chromatic, quasi-randomly disordered lattice potential. We identify the many-body localized phase by monitoring the time evolution of local observables following a quench of system parameters. Specifically, we prepare a high-energy initial state with strong charge density wave (CDW) order (as shown in Fig.\ 1A) and measure the relaxation of this charge density wave in the ensuing unitary evolution.  Our main observable is the imbalance $\mathcal{I}$ between the respective atom numbers on even ($N_e$) and odd ($N_o$) sites

\begin{equation}
	\mathcal{I} = \frac{N_e-N_o}{N_e+N_o}, 
	\label{Imbalance_equn}
\end{equation}
which directly measures the CDW order.  While the initial CDW ($\mathcal{I} \gtrsim 0.9$) will quickly relax to zero in the thermalizing case, this is not true in a localized system, where ergodicity is broken and the system cannot act as its own heat bath (Fig.\ 1B) \cite{Iyer13}.  Intuitively, if the system is strongly localized, all particles will stay close to their original positions during time evolution, thus only smearing out the CDW a little.  A longer localization length $\xi$ corresponds to more extended states and will lead to a lower steady state value of the CDW. The long-time stationary value thus effectively serves as an order parameter of the MBL phase and allows us to map the phase boundary between the ergodic and non-ergodic phases in the parameter space of interaction versus disorder strength.  In particular, in the non-interacting system the CDW vanishes asymptotically as $\propto 1/\xi^2$ \cite {SOMs}.  In contrast to previous experiments, which studied the effect of disorder on the global expansion dynamics \cite{Billy08,Roati08,Deissler10,Kondov2013,Derrico2014}, the CDW order parameter acts as a purely local probe, directly capturing the ergodicity breaking.  

%In a non-interacting system we expect the stationary value to behave as $1/\left(1+\xi^2\right)$, with $\xi$ being the localization length \cite{SOMs}.   
 
%Our approach of targeting the relaxation of local observables should be contrasted with measurements of center of mass motion, previously used to investigate many-body localization in an ultra cold Fermi gas \cite{Kondov2013}. One advantage of our approach is that it is insensitive to the large scale inhomogeneity induced by the trap. In mass transport measurements by contrast, a more localized region found at the edges of the trap can have an overwhelming effect by blocking the core of the cloud. Furthermore the saturation value of the density wave provides quantitative information on the degree of localization rather than just an indication that the system is localized. Finally, even more information can be gleaned from analyzing the full relaxation dynamics of local observables \cite{Andraschko2014,Vasseur2014,Serbyn2014a}.

Our system can be described by the one-dimensional fermionic Aubry-Andr\'{e} model \cite{Aubry80} with interactions \cite{Iyer13}, given by the Hamiltonian

\begin{equation}
\begin{split}
	\hat{H} = &-J \sum_{i,\sigma} \left(\hat{c}_{i,\sigma}^{\dagger} \hat{c}_{i+1,\sigma}+\text{h.c.}\right) \\ &+\Delta\sum_{i,\sigma} \cos (2\pi\beta i+\phi)\hat{c}_{i,\sigma}^{\dagger}\hat{c}_{i,\sigma} +U\sum_i \hat{n}_{i,\uparrow}\hat{n}_{i,\downarrow}.
	\label{AA_hamiltonian}
\end{split}	
\end{equation}
Here, $J$ is the tunneling matrix element between neighboring lattice sites and $\hat{c}_{i,\sigma}^{\dagger}$ ($\hat{c}_{i,\sigma}$) denotes the creation (annihilation) operator for a fermion in spin state $\sigma\in\{\uparrow,\downarrow\}$ on site $i$.   The second term describes the quasi-random disorder, i.e.\ the shift of the on-site energy due to an additional incommensurate lattice, characterized by the ratio of lattice periodicities $\beta$, disorder strength $\Delta$ and phase offset $\phi$.  Lastly, $U$ represents the on-site interaction energy and $\hat{n}_{i,\sigma} = \hat{c}_{i,\sigma}^{\dagger}\hat{c}_{i,\sigma}$ is the local number operator (see Fig.\ 1C).  

\begin{Figure}%[h!t]
	\centering
	\includegraphics[width=84mm]{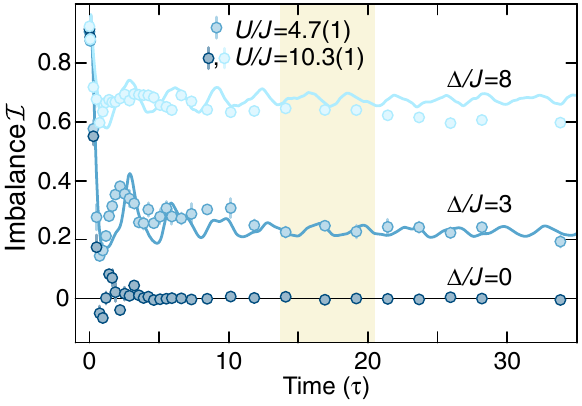} %76 %was 84mm   58mm
	\captionof{figure}{{\small{\textbf{Time evolution of an initial charge-density wave.} A charge density wave, consisting of fermionic atoms occupying only even sites, is allowed to evolve in a lattice with an additional quasi-random disorder potential. After variable times the imbalance $\mathcal{I}$ between atoms on odd and even sites is measured. Experimental time traces (circles) and DMRG calculations for a single homogeneous tube (lines) are shown for various disorder strengths $\Delta$. Each experimental datapoint denotes the average of six different realizations of the disorder potential and the error bars show the standard deviation of the mean. The shaded region indicates the time window used to characterise the stationary imbalance in the rest of the analysis.}}}
	\label{time_traces}
\end{Figure}

\begin{Figure}%[h!t]
	\centering
	\includegraphics[width=84mm]{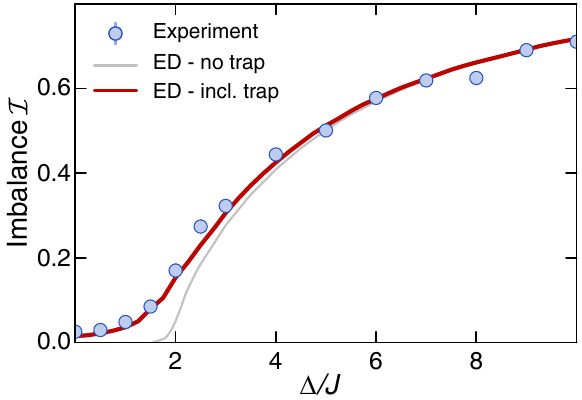} %76 %was 84mm  \linewidth  58mm
	\captionof{figure}{{\small{\textbf{Stationary values of the imbalance $\mathcal{I}$ as a function of disorder $\Delta$ for non-interacting atoms.}  The Aubry-Andr\'{e} transition is at $\Delta/J = 2$. Circles show the experimental data, along with Exact Diagonalization (ED) calculations with (red line) and without (grey line) trap effects. Each experimental data point is the average of three different evolution times (13.7$\tau$, 17.1$\tau$ and 20.5$\tau$) and four different disorder phases $\phi$, for a total of 12 individual measurements per point.  To avoid any interaction effects, only a single spin component was used. The ED calculations are averaged over similar evolutions times to the experiment and 12 different phase realizations. Error bars show the standard deviation of the mean.}}}
	\label{non_interacting}
\end{Figure} 

\begin{figure*}[t!]
	\centering
	\includegraphics[width=180mm]{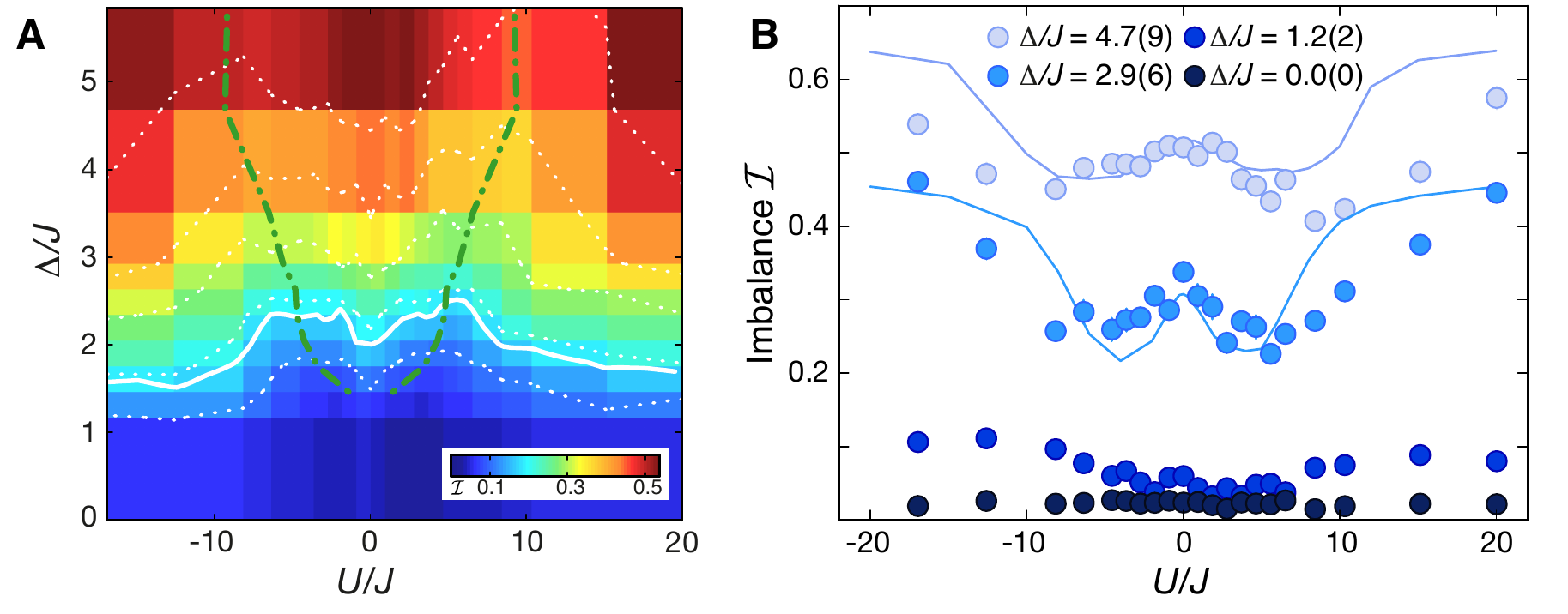} %76 %was 84mm
	\caption{{\small{\textbf{Stationary imbalance for various interaction and disorder strengths.} \textbf{A}: Stationary Imbalance $\mathcal{I}$ as a function of interactions $U$ and disorder strength $\Delta$. Moderate interactions reduce the degree of localization compared to the non-interacting or strongly interacting cases.  The white dotted lines are contours of equal $\mathcal{I}$, while the solid white line is the contour of $\mathcal{I}$ matching the Aubry-Andr\'{e} transition ($U=0$ and $\Delta/J= 2$) extended to the interacting case. It indicates the MBL transition. The green dot-dashed line shows the fitted minima of $\mathcal{I}$ for each $\Delta$ \cite{SOMs}. Each individual data point (vertices of the pseudo-color plot) is the average of the same 12 parameters as in Fig.\ 3. The color of each square represents the average imbalance of the four points on the corners.  All data taken with a doublon fraction of $\simeq 34\left(2\right) \,\%$. \textbf{B}:  Cuts along four different disorder strengths. The effect of interactions on the localization gives rise to a characteristic 'W'-shape. Solid lines are the results of DMRG simulations for a single homogeneous tube.  Error bars indicate the standard deviation of the mean.}  }}
	\label{2D_MBL_map}
\end{figure*}

This quasi-random model is special in that, for almost all irrational $\beta$ \cite{SOMs}, all single particle states become localized at the same critical disorder strength $\Delta/J=2$ \cite{Aubry80}.  For larger disorder strengths the localization length decreases monotonically. Such a transition was indeed observed  experimentally in a non-interacting bosonic gas \cite{Roati08}. In contrast, truly random disorder will lead to single-particle localization in one dimension already for arbitrarily small disorder strengths. Previous numerical work indicates many-body localization in quasi-random systems to be similar to that obtained for a truly random potential \cite{Iyer13}.  

%Note that localization persists for all interaction strengths -- even those larger than the disorder strength (i.e. for $U \gtrsim \Delta$).

\paragraph*{Experiment}

We experimentally realize the Aubry-Andr\'{e} model by superimposing on the primary, short lattice ($\lambda_{s}=532\,\textrm{nm}$) a second, incommensurate disorder lattice with $\lambda_{d}=738\,\textrm{nm}$ (thus $\beta = \lambda_s/\lambda_d \approx 0.721$) and control $J$, $\Delta$ and $\phi$ via lattice depths and relative phase between the two lattices \cite{SOMs}.  The interactions ($U$) between atoms in the two different spin states $\left|\uparrow\right\rangle$ and $\left|\downarrow\right\rangle$ are tuned via a magnetic Feshbach resonance \cite{SOMs}.  In total, this provides independent control of $U$, $J$ and $\Delta$ and enables us to continuously tune the system from an Anderson insulator in the non-interacting case to the MBL regime for interacting particles.

%For non-interacting atoms ($U = 0$), Anderson localisation \cite{Anderson58} is predicted to occur in the Aubry-Andr\'{e} model above a critical disorder strength of $\Delta/J = 2$ \cite{Aubry80}, as was observed in \cite{Roati08,Deissler10}.  While the Aubry-Andr\'{e} model only has quasi-random disorder, based on previous numerical work we expect the generic many-body localisation in this system to be similar to that obtained for a usual random potential \cite{Iyer13}.  This means that the Aubry-Andr\'{e} model is in some respects superior for studying MBL, as the transition is at a non-zero disorder strength.  In contrast, for Anderson localisation any finite disorder will lead to localisation \cite{Anderson58}.The green dot-dashed line shows the minimum value of $\mathcal{I}$, fitted as described in \cite{SOMs}.

An additional long lattice ($\lambda_{l}=1064\textrm{nm} = 2\lambda_{s}$) forms a period-two superlattice \cite{SebbyStrabley06,Foelling07} together with the short lattice and is employed during the preparation of the initial CDW state, and during detection \cite{SOMs}.  Deep lattices  along the orthogonal directions ($\lambda_{\perp} = 738\textrm{nm}$ and $V_{\perp}=36(1)E_R$), create an array of decoupled 1D tubes. Here, $E_R=h^2/\left(2m\lambda_{\textrm{lat}}^2\right)$ denotes the recoil energy, with $h$ being Planck's constant, $m$ the mass of the atoms and $\lambda_{\textrm{lat}}$ the respective wavelength of the lattice lasers. 

We employ a two component degenerate Fermi gas of $^{40}$K atoms, consisting of an equal mixture of 25-30 $\times 10^3$ atoms in each of the two lowest hyperfine states $\left|F,m_F\right\rangle = \left|\frac{9}{2},-\frac{9}{2}\right\rangle \equiv \left|\downarrow\right\rangle$ and $\left|\frac{9}{2},-\frac{7}{2}\right\rangle \equiv \left|\uparrow\right\rangle$, at an initial temperature of 0.24(2) $T_F$, where $T_F$ is the Fermi temperature. The atoms are initially prepared in a finite temperature band insulating state \cite{Schneider08} in the long and orthogonal lattices. We then split each lattice site by ramping up the short lattice in a tilted configuration \cite{SOMs} and subsequently ramp down the long lattice. This creates a charge density wave, where there are no atoms on odd lattice sites but zero, one or two atoms on each even site \cite{Foelling07,Trotzky12}.  This initial CDW is then allowed to evolve for a given time in the 8.0(2)$E_R$ deep short lattice at a specific interaction strength $U$ in the presence of disorder $\Delta$. In a final step, we detect the number of atoms on even and odd lattice sites by employing a band-mapping technique which maps them to different bands of the superlattice \cite{Trotzky12,SOMs}.  This allows us to directly measure the imbalance $\mathcal{I}$, as defined in Eq. (\ref{Imbalance_equn}).

%This state is then allowed to evolve for a given time in the 8.0(2)$E_R$ short lattice (which is deep in the tight binding limit) at a given $U$ and $\Delta$, after which we measure the number of atoms on even ($n_e$) and odd ($n_o$) lattice sites using the superlattice \cite{Trotzky12} \cite{SOMs} and extract the imbalance $\mathcal{I}$ between the respective atom numbers 

\paragraph*{Results}  

We track the time evolution of the imbalance $\mathcal{I}$ for various interactions $U$ and disorder strengths $\Delta$ (see Fig.\ 2).
At short times the imbalance exhibits some dynamics consisting of a fast decay followed by a few damped oscillations.  After a few tunneling times $\tau = h/(2\pi J)$ the imbalance approaches a stationary value.  In a clean system ($\Delta/J = 0$) and for weak disorder, the stationary value of the imbalance approaches zero.  For stronger disorder, however, this behaviour changes dramatically and the imbalance attains a non-vanishing stationary value that persists for all observation times.  Since the imbalance must decay to zero on approaching thermal equilibrium at these high energies, the non-vanishing stationary value of $\mathcal{I}$ directly indicates non-ergodic dynamics.  Deep in the localized phase, where unbiased numerical Density-Matrix Renormalisation Group (DMRG) calculations are feasible due to the slow entanglement growth, we find the stationary value obtained in the simulations to be in very good agreement with the experimental result.
These simulations were performed for a single homogeneous tube without any trapping potentials \cite{SOMs}. The stronger damping of oscillations observed in the experiment can be attributed to a dephasing caused by variations in $J$ between different 1D tubes \cite{Trotzky12,SOMs}. 
 
We experimentally observe an additional very slow decay of $\mathcal{I}$ on a timescale of several hundred tunneling times  for all interaction strengths, which we attribute to the fact that our system is not perfectly closed due to small losses, technical heating and photon scattering \cite{Pichler10,SOMs}. Another potential mechanism for delocalization at long times is related to the intrinsic SU(2) spin symmetry in our system \cite{arXiv_1410_6165}. However, for the relevant observation times our numerical simulations do not indicate the presence of such a thermalization process.

%Local clusters of fermions that accidentally couple to a large (nearly classical) spin can serve as nucleation sites for thermalization . However, since we do not see any sign of such a delocalization in the DMRG calculations, we conclude that this mechanism, if at all effective, is sub-dominant compared to the above sources.

To characterize the dependence of the localization transition on $U$ and $\Delta$, we focus on the stationary value of $\mathcal{I}$, plotted in Fig.\ 3 for non-interacting atoms and in Fig.\ 4 for interacting atoms.  For non-interacting atoms (Fig.\ 3), the measured imbalance is compatible with extended states within the finite, trapped system for $\Delta/J \lesssim 2$.  Above the critical point of the homogeneous Aubry-Andr\'{e} model at $\Delta/J = 2$ \cite{Aubry80}, however, the measured imbalance strongly increases as the single-particle eigenstates become more and more localized.  The observed transition agrees well with our theoretical modeling including the harmonic trap  \cite{SOMs}.

%The stationary value of $\mathcal{I}$ depends on $U/J$ and $\Delta/J$: for weak disorders ($\Delta/J \lesssim 2$), the equilibrium imbalance is zero, as expected for this thermalizing system.  For stronger disorder ($\Delta/J > 2$),    Theoretical simulations are also shown, which exhibit more slowly decaying oscillations.  This is due to slight shifts in $J$ and $\Delta$ across the experimental atomic cloud due to the gaussian shape of the laser beams \cite{Trotzky12,Pertot14,SOMs} (FLESH OUT ONCE WE FINALISE WHICH THEORY CURVES TO INCLUDE).

%variations in $\Delta/J$ across the cloud \cite{SOMs}.

%The first localisation transition we investigate is the non-interacting ($U = 0$) case, where there should be a transition from delocalised to localised at $\Delta = 2$ (the Aubry-Andr\'{e} critical point \cite{Aubry80}).  Fig.3A shows our experimental data, plotted alongside the theoretical values for a single particle.  To ensure that the atoms really are non-interacting, we only use the $\left|\downarrow\right\rangle$ spin component (see SOMs for details), such that the Pauli exclusion principle ensures there can be no interactions between the individual fermions.  A transition is observed slightly below $\Delta = 2$, in agreement with our theoretical modelling (shown as grey solid line).  The shift from the ideal theoretical case of $\Delta = 2$ is due to a number of factors such as the dipole trap and variations in $\Delta/J$ across the cloud (see SOMs for details).  These factors also lead to a slight broadening of the transition.

\begin{Figure}%[h!t]
	\centering
	\includegraphics[width=84mm]{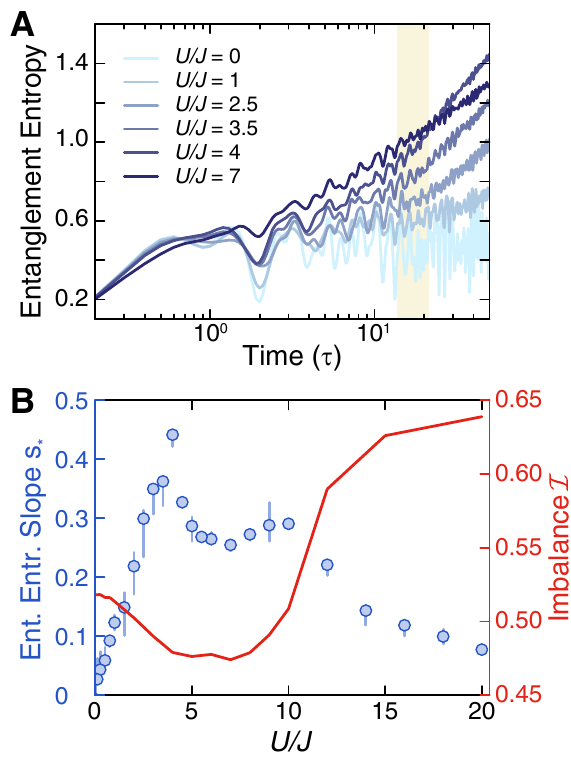} %76 %was 84mm    58mm
	\captionof{figure}{{\small{\textbf{Growth of entanglement entropy and corresponding slope.} \textbf{A:} DMRG results of the entanglement entropy growth for various interaction strengths and $\Delta/J=5$.  For long times, logarithmic growth characteristic of interacting MBL states is visible.  The experimentally used evolution times indicated by the yellow shaded region are found to be in the region of logarithmic growth. \textbf{B:} The slope of the logarithmic growth, extracted using linear fits up to the longest simulated time ($50\,\tau$) in A, shows a non-monotonic dependence on the interaction strength, which tracks the inverse of the steady state CDW value (red line). Error bars reflect different initial starting times for the fit.}}}
	\label{entanglement_entropy}
\end{Figure}

The addition of moderate interactions slightly reduces the degree of localization compared to the non-interacting case, i.e.\ they decrease the imbalance $\mathcal{I}$ and hence increase the critical value of $\Delta$ necessary to cross the delocalization-localization transition (Fig.\ 4A and B).  Importantly, we find that localization persists for all interaction strengths. For a given disorder, the imbalance $\mathcal{I}$ decreases up to a value of $U \sim 2\Delta$ before increasing again. For large $|U|$, the system even becomes more localized than in the non-interacting case.  This can be understood qualitatively by considering an initial state consisting purely of empty sites and sites with two atoms (doublons): for sufficiently strong interactions, isolated doublons represent stable quasiparticles as the two atoms cannot separate and hence only tunnel with an effective second-order tunneling rate of $J_D=\frac{2J^2}{|U|} \ll J$ \cite{Winkler06,Trotzky08}.  This strongly increases the effective disorder $\propto \Delta/J_D \gg \Delta/J$ and promotes localization. In the experiment, the initial doublon fraction is well below one \cite{SOMs} and the density is finite, such that we observe a weaker effect.  We find the  localization dynamics and the resulting stationary values to be symmetric around $U=0$, highlighting the dynamical $U \leftrightarrow -U$ symmetry of the Hubbard Hamiltonian for initially localized atoms \cite{Schneider12}.  The effect of interactions can be seen in the contour lines (Fig.\ 4A, dotted white lines) as well as directly in the characteristic `W' shape of the imbalance at constant disorder (Fig.\ 4B), demonstrating the re-entrant behaviour of the system \cite{Lev14}.  This behaviour extends to our best estimate of the localization transition, which is shown in Fig.\ 4A as the solid white line.  

%This behaviour extends to our best estimate of the localization transition (solid white line), namely the imbalance corresponding to $\Delta/J = 2$ in the non-interacting case.

%In order to determine the interaction dependence of the transition point from many-body ergodic to localised, we additionally plot a contour line (white) for the imbalance value measured at the known critical point of $\Delta/J = 2$ of the non-interacting model.  This contour line also illustrates the characteristic `W' structure in the data, demonstrating the re-entrant behaviour of the system.  %The error bars for this curve are taken from contours plotted at the extent of the error bars from the value of $\mathcal{I}_c$, and are shown in Fig. 3C as the faded grey area.

\begin{Figure}%[h!t]
	\centering
	\includegraphics[width=84mm]{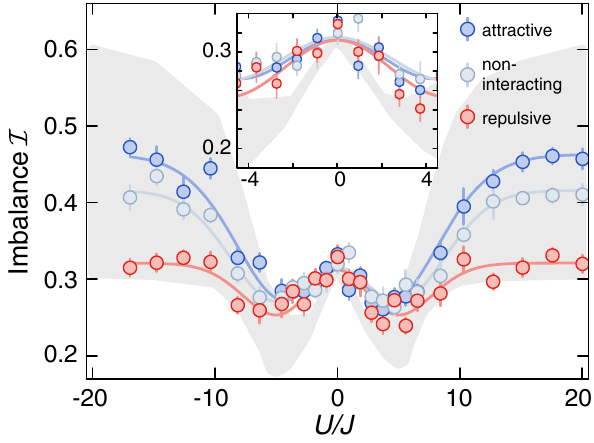} %76 %was 84mm    58mm
	\captionof{figure}{{\small{\textbf{Stationary imbalance $\mathcal{I}$ as a function of interaction strength during loading.} Data taken with disorder $\Delta/ J = 3$.  The loading interactions of $a_{\text{load}} = -89(2)\,\textrm{a}_0$ (attractive, where $\textrm{a}_0$ denotes Bohr's radius), $0(1)\,\textrm{a}_0$ (non-interacting) and $142(1)\,\textrm{a}_0$ (repulsive) correspond to initial doublon fractions of 51(1)\%, 43(2)\%, and 8(6)\%, respectively \cite{SOMs}.   Each $\mathcal{I}$ value is the average of the same 12 parameters as in Fig.\ 3.  Error bars show the standard deviation of the mean.  Solid lines are guides to the eye.  The grey shaded area spans the limiting cases of 0 and 50\% doublons, simulated using DMRG for a single homogeneous} tube.}}
	\label{D3_load_U}
\end{Figure}

We can gain additional insight into how localization changes with interaction strength by computing the growth of the entanglement entropy between the two halves of the system during the dynamics, as shown in Fig.\ 5A. For long times, we observe a logarithmic growth of the entanglement entropy with time as $S(t) = S_{{\rm offset}} + s_{*} \ln(t/\tau)$, which is characteristic of the MBL phase \cite{Znidaric2008,Bardarson2012}. The slope $s_{*}$ is proportional to the bare localization length $\xi_*$, which in a weakly interacting system in the localized phase corresponds to the single particle localization length. In general, $\xi_*$ is the characteristic length over which the effective interactions between the conserved local densities decay \cite{Serbyn2013a,Huse2013a} and connects to the many-body localization length $\xi$ deep in the localized phase. In contrast to $\xi$, however, $\xi_{*}$ is expected to remain finite at the transition \cite{Vosk2014-2}. We find $s_{*}$ to exhibit a broad maximum for intermediate interaction strengths (Fig. 5B), corresponding to a maximum in the thus inferred localization length.  This maximum in turn leads to a minimum in the CDW value.  The characteristic `W' shape in the imbalance is thus directly connected to the maximum in the entanglement entropy slope, as both are consequences of the maximum in localization length.  Equivalent information on the localization properties as obtained from the entanglement entropy can be gained in experiments by monitoring the temporal decay of fluctuations around the stationary value of the CDW \cite{SOMs}.  While we do not have sufficient sensitivity to measure these fluctuations in the current experiment, we expect them to be accessible to experiments with single site resolution \cite{Bakr09,Sherson10}.

To systematically study the effect of the initial energy density on the MBL phase, we load the lattice using either attractive, vanishing or repulsive interactions (Fig.\ 6), thereby predominantly changing the number of doublons in the initial state \cite{SOMs}. Since the initial state consists of fully localized particles only, the local energy density is directly given by the product of interaction strength $U$ and doublon density. We find that for an interaction strength during the evolution of $|U/J|\leq 6$ the energy density does not significantly affect the localization properties, proving that MBL persists over a wide energy range. For $|U/J| > 8$, localization properties depend significantly on the doublon fraction because of the second emerging energy scale $J_{D}$, as discussed above. For the case of repulsive loading, i.e.\ a low fraction of doubly occupied sites, the imbalance for $U/J=0$ and strong interactions match within error.  Indeed, a rigorous mapping can be made between the non-interacting system and the dynamics in the doublon free subspace at strong interactions $|U/J| \to \infty$ \cite{SOMs}. At very large interactions and high doublon fractions, the additional long timescales start to also compete with heating and loss processes, rendering the definition of stationary states challenging. %showing that in the absence of additional doublon physics the non- and strongly interacting regimes are equivalent.
%{\large\textbf{Conclusion}}

\paragraph*{Conclusion}

We have created and characterized the many-body localized phase of a system of interacting ultracold fermions in a quasi-random optical lattice.  The dependence of the MBL phase-transition point on interactions was measured by studying the evolution of an artificially created charge density wave.  Moderate interactions have a delocalizing effect and increase the critical disorder strength. We have also shown the MBL phase to be stable over a wide range of energy densities.

Our experimental demonstration of ergodicity breaking due to many-body localization paves the way for many further investigations.  An interesting extension would be to use `true' random disorder created by e.g.\ an optical speckle pattern, as has been used to study Anderson localization \cite{Billy08}.  Another important next step is extending the present study to higher dimensions.  Additional insight can also be gained by analyzing the full relaxation dynamics of local observables \cite{Andraschko2014,Vasseur2014,Serbyn2014a}, in an experimental setup featuring single site resolution \cite{Bakr09,Sherson10}. For instance, the decay of fluctuations of $\mathcal{I}$ with time could be directly measured, providing an even more direct connection to the entanglement entropy. Another important direction for future investigation is the effect of opening the system in a controlled way. This could be done, e.g.\ by adding a near-resonant laser to deliberately enhance photon scattering or by employing a Bose-Fermi mixture, in which excitations of the BEC form a well controlled bath for the fermions. This will allow a systematic study of the critical dynamics associated with the MBL phase transition, where the bath relaxation time now provides the only scale. Such a study would also allow the MBL phase to be clearly distinguished from classical glassy dynamics. The latter, unlike MBL, is insensitive to coupling of the system to an external bath.

%A different complex area is MBL in a Bose-Fermi mixture where excitations of the BEC form a well controlled bath for the fermions.  Another challenging question will be the fate of many-body localization for systems coupled to thermal environment and the connection to glass-like dynamics [References?]. Further investigations into MBL will also provide insights into more broader questions regarding quantum thermalization, as it is an example of a system where the eigenstate thermalization hypothesis \cite{Rigol08} is false \cite{Iyer13}.  Finally, as it occurs at high-energy, MBL offers the possibility to study a unique type of quantum phase transition. 

%\printbibliography[segment=\therefsegment,check=onlynew,heading=subbibliography]

%\bibliographystyle{Science}  %ieeetr
%\bibliography{negative_T_refs}

\vspace{0.0cm}

\noindent
\textbf{Acknowledgments:} We acknowledge technical assistance by D. Garbe and F. G\"org during the setup of the experiment. We acknowledge financial support by the Deutsche Forschungsgemeinschaft (FOR801, Deutsch-Israelisches Kooperationsprojekt Quantum phases of ultracold atoms in optical lattices), the European Commision (UQUAM, AQuS), the US Defense Advanced Research Projects Agency (Quantum Emulations of New Materials Using Ultracold Atoms), the Minerva Foundation, ISF grant \# 1594/11 and Nanosystems Initiative Munich (NIM).

\vspace{0.4cm}

%\end{multicols}

\newpage

\begin{figure*}[h!t]
	\centering
	\includegraphics[width=180mm]{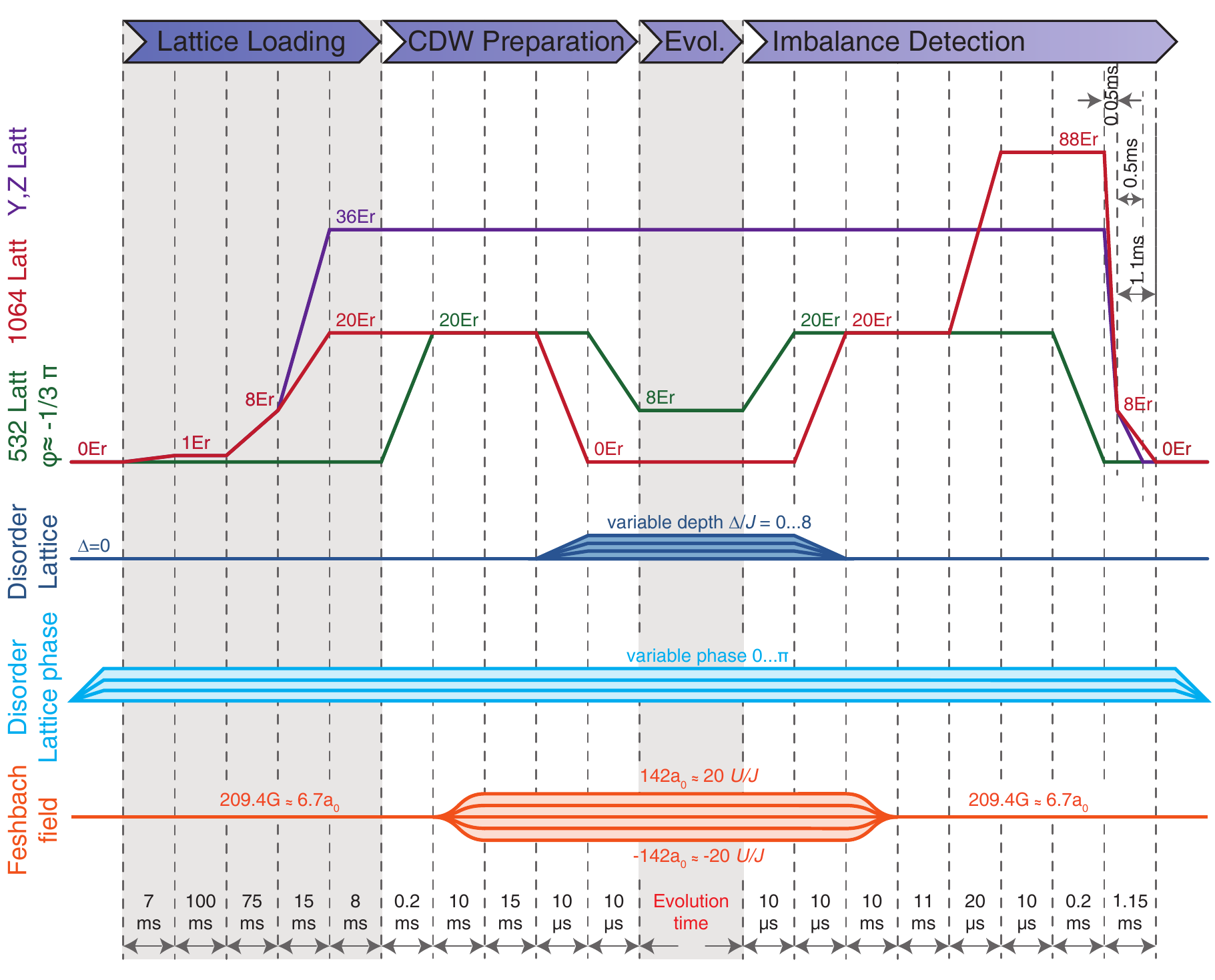} %76 %was 84mm
	\caption{{\small{\textbf{Full experimental  sequence.}  Schematic showing lattice depths, phase and Feshbach field ramps for loading, CDW preparation, evolution and detection of the imbalance.  See text for details.}}}
	\label{sequence}
\end{figure*}

\newpage

\noindent
\textsf{\textbf{Supplementary Material:}}

\noindent
\textsf{Supplementary Text}

\setcounter{section}{0}

\paragraph*{General sequence}

The experimental sequence for producing degenerate Fermi gases of $^{40}$K  via sympathetic cooling with $^{87}$Rb  in a magnetic quadrupole and optical dipole trap followed by evaporative cooling has been described elsewhere \cite{Schneider08}. Final clouds contain typically 50-60k $^{40}$K atoms at a temperature of 0.24(2) times the Fermi temperature $T_F$, with 50(3)\% of atoms in each of the $\left|\uparrow\right\rangle$ and $\left|\downarrow\right\rangle$ spin states.  We tune the interactions between the two spin states via a Feshbach resonance  at 202.1G \cite{Regal03}.  All experiments described in this work employ absorption imaging after $8\,\textrm{ms}$ time of flight (TOF) along the $y$ axis ($z$ axis is oriented along the vertical direction).  This is orthogonal to the $x$ axis along which the superlattice is aligned and all experiments take place. For the experiments shown in Fig.\ 3, where we employ a single spin state, we set the magnetic field early in the evaporation sequence to $469\,\textrm{G}$, which is the centre of a Feshbach resonance between $\left|\uparrow\right\rangle$ $^{40}$K atoms and $^{87}$Rb atoms in the $\left|F=2,m_F=2\right\rangle$ state \cite{Klempt07}.  This results in all $\left|\uparrow\right\rangle$ atoms being lost (along with some $^{87}$Rb atoms).  By adjusting the cooling sequence slightly we are able to prepare a sample with 98(5)\% of atoms in the $\left|\downarrow\right\rangle$ state with the same temperature and atom number as in the usual sequence.

\paragraph*{Lattice Details:}

\subparagraph*{Lattice setup}

All lattice potentials result from standing waves created by retro-reflected single-frequency laser beams along the $x$, $y$ and $z$ axes.  All beams have a Gaussian profile and are focussed down to $1/e^2$ waists of $\sim 150 \, \mu \textrm{m}$ at the positions of the atoms. As our cloud has a FWHM of $\sim 50\,\mu\textrm{m}$ in the horizontal plane and $\sim 15\,\mu\textrm{m}$ in the vertical direction, this results in slight variations of the lattice parameters $J$, $U$ and $\Delta$ across the cloud (see below).

\subparagraph*{Superlattice lock}

In our superlattice setup the $1064\,\textrm{nm}$ laser (long lattice) serves as a master oscillator and its frequency is locked to a thermally stabilized and acoustically isolated Fabry-P\'erot cavity. The absolute short term stability of the master oscillator is $65\,\textrm{kHz}$ over $100\,\textrm{ms}$ and is characterized through the residual locking error. The short lattice laser at $532\, \textrm{nm}$ is offset locked relative to the second harmonic of the long lattice laser. The offset lock frequency sets the relative phase of the superlattice potential \cite{Foelling07}. We obtain a relative line width of the offset lock of $1.6\,\textrm{MHz}$, which corresponds to a phase fluctuation of $\delta\phi=14\,\textrm{mrad}$.  This also dominates the  absolute short term stability of the short lattice laser.

\subparagraph*{Phase adjustment of the superlattice}

In order to calibrate the phase of the superlattice, we prepare a spinpolarized band insulating cloud in a combination of long and perpendicular lattices and subsequently split each well non-adiabatically by adding the short lattice. Time of flight pictures of such arrays of double wells show the typical double-slit interference pattern only for phase settings of the superlattice that result in a symmetric splitting. By determining the difference in offset frequencies necessary to reach neighboring symmetric configurations, we can precisely calibrate the phase $\phi$ of the superlattice.

\subparagraph*{Disorder lattice}

We use a Coherent MBR-110  Ti:Sa laser operated at $738\,\textrm{nm}$ to create the disorder potential and the perpendicular lattices. This laser is stabilized via its internal reference cavity, which can be externally tuned to change the laser frequency. An additional external temperature stabilized and vibration isolated Fabry-P\'erot cavity is used to measure the linewidth and to calibrate the frequency changes necessary to set the phase $\phi$ of the disorder lattice. The phase  $\phi$ is always set before the lattice sequence starts and kept constant throughout the experiment.

Note that, strictly speaking, the incommensurate ratio $\beta \approx 532 / 738$ should be an irrational Diophantine number \cite{Jitomirskaya99} for there to be a sharp localization transition in the Aubry-Andr\'e model.  However, for a realistic finite system, it is sufficient for the actual period of the combined disorder and regular lattices to exceed the system size \cite{Modugno09}.

\subparagraph*{Disorder strength}

The disorder strength $\Delta$ in the tight-binding limit depends on the lattice depth of the primary lattice  $V_{s}=s_{s}\cdot E_{R, 532}$, the disorder lattice depth $V_{d}=s_{d}\cdot E_{R,738}$ and the  wavelength ratio $\beta \approx 532 / 738$.  It is given by \cite{Modugno09}     

\begin{equation}
	\begin{split}
	\Delta = &\frac{s_{d} \beta^{2}}{2} k_s E_{R,532} \int dx \cos(2\beta k_s x \left|w\left( k_s x \right)\right|^{2}) 
	%\\= &\beta^2 \cdot s_{d} \cdot O  
	%\\ \approx  &0.52 \cdot s_{d} \cdot 0.37
	\\ \approx &0.2 \cdot s_{d} \cdot E_{R,532} 
	\end{split}
	\label{Delta_def}
\end{equation}
where %$s_{Disorder}$ is the lattice depth of the disorder lattice in $ E_{R,738} = \frac{\hbar^2 k_{738}^2}{2m}$,$alpha=\frac{\lambda_{short}}{\lambda_{disorder}}$ the (incommensurate) ratio of the lattice wavelengths and 
$w\left( x \right)$ denotes the Wannier function of the unperturbed short lattice, $k_s = 2\pi/\lambda_s$, and  $s_s$ and $s_d$ are the depths in terms of $E_R$ of the short and disorder lattices, respectively. Here,  $\Delta$ is given in units of $E_{R,532}=h \times 17.64\,\textrm{kHz}$. We use a primary lattice depth of $s_s=8$ yielding an undisturbed tunnel coupling of $J= h \times 540\,\textrm{Hz}$. Combining these numbers, we obtain a conversion factor for 

\begin{equation}
\frac{\Delta}{J}=\frac{0.2 \cdot s_{d} \cdot E_{R,532}}{h \cdot 550\,\textrm{Hz}}\approx 6.67 \cdot s_{d}.
\end{equation}

\paragraph*{Long-Term evolution}

For extremely long hold times ($t\gg100\,$ms or $500\tau$) we observe an additional decay of the imbalance, along with the total particle number. This decay can be attributed
to unwanted additional effects, such as off-resonant light scattering and collisions with background gas atoms, as well as technical imperfections including laser noise, frequency instabilities or vibrations and other pointing instabilities. All of these effects violate the condition of a perfectly closed system and will therefore lead to a vanishing imbalance in the long-time limit. Since the relevant timescale is, however, more than an order of magnitude larger than the longest times used in the experiment, these effects have a negligible impact on the reported observations. 

%For extremely long time scales, we start to see effects due to the technical limitations of the experiment.  In the standard lattice used in these experiments, the lattice lifetime over which we see heating and atom losses is $\sim$ 800ms (= 2800$\tau$), and is essentially independent of parameters such as $\Delta$ and $U$.  As this is over two of orders of magnitude larger than the timescales used in this experiment, it will have almost no impact.  We attribute this lifetime to technical issues such as off-resonant light scattering, laser beam jitter and frequency instability. 

%800(20)ms = 2800(80) tau lattice lifetime

%Additionally, we observe a decay in $\mathcal{I}$ at long times, which for the non-interacting case is on a comparable timescale to the lattice lifetime.  For $|U|>0$ this decay is faster, which we believe is due to losses perturbing the many-body state, which promotes tunneling.  However, since this timescale is always $>500\tau$, it will have a negligible impact on our observations, which are taken on significantly shorter timescales. 

\paragraph*{Trap effects on non-interacting dynamics} 

Compared to the homogeneous Aubry-Andr\'{e} model, the experimental results 
are modified due to the presence of the 3D harmonic trap.  All experimental results are automatically averaged over many inequivalent 1D tubes and, furthermore, the atoms in each
tube experience a confining harmonic potential. 
 
In the presence of a longitudinal harmonic trapping potential (trapping frequency $\omega_x\approx2\pi\times 60\,$Hz), the sharp transition at $\Delta/J = 2$ expected in the homogeneous case \cite{Aubry80} becomes smeared out (see Fig.\ 3), as even in a weak trap the localization length is bounded from above by the finite cloud size. In a sufficiently strong trap, particles situated far away from the trap center will experience offset energies that are bigger than $4J$ and thereby become localized to their respective side of the trap, even in the absence of any disorder. Nonetheless, although the trap causes some degree of localization even in the absence of disorder, we still observe a clear crossover from trap dominated behavior to disorder-induced localization.

The finite-size of the used lattice beams (waists $w\approx150\,\mu\textrm{m}$) results in  an additional small but finite variation of $J$ between the central and off-center tubes which increases with the distance from the beam center. More than $90\%$ of the atoms see a less than $20\%$ change in the hopping term. The Rayleigh length of the lattice beams is long enough that there is no appreciable change in tunneling along the tubes. The above variation in $J$ mainly gives rise to a dephasing between the dynamics in individual tubes, and is responsible for the more pronounced damping of imbalance oscillations compared to the homogeneous case (see Fig.\ \ref{time_trace_non_int}). In the ideal homogeneous system with zero disorder and no variation in $J$, the imbalance would decay according to a Bessel function \cite{Pertot14}. 
 
Lastly, since the quasi-random potential is produced by superimposing a disorder lattice on top of the primary lattice, this additional potential gives rise to a longitudinal modulation of the hopping rate. This can be modeled as a sinusoidal variation of the same periodicity as the quasi-random potential (i.e.\ as $\text{sin}(2\pi\beta i + \phi + \phi_{\text{diff}})$, where $\phi_{\text{diff}}$ is the phase difference between the modulation of on-site energy and tunneling. For $\Delta/J=3$, the amplitude of this modulation is about $8\%$ of the hopping term. This modulation does not change the average hopping in a tube and hence does not change the average imbalance. However, it further increases the damping of the imbalance oscillations.

\paragraph*{CDW preparation sequence}

The preparation sequence is illustrated in Fig.\ \ref{sequence}.  To prepare the initial CDW state we start with a 3D band insulating state in a $20\,E_R$ deep long lattice and $36\, E_R$  deep orthogonal lattices.  The short lattice is then ramped up in $200\,\mu s$ to $20\, E_R$ with the superlattice phase set to $\phi \sim 3\pi/4$.  This creates an array of tilted double wells, where the tilt offsets are large enough for all atoms to be loaded into the even wells.  This tilt additionally suppresses all tunnelling while we set the interaction strength $U$ by ramping the Feshbach field to the desired value in $25\,\textrm{ms}$.  The long lattice is then switched off in $10\, \mu \textrm{s}$ while the disorder lattice is simultaneously ramped up.  Finally, the short lattice is ramped down to $8 E_R$ in $10\, \mu \textrm{s}$, thereby enabling tunneling and initiating the dynamics. The transverse lattices remain deep to decouple individual tubes.  

This preparation sequence yields 96(2)\% of atoms on even sites in the short lattice.  We attribute the small residual number of atoms sitting on odd sites to non-adiabaticities in the splitting process, tunneling events during the preparation sequence, higher band occupancies and detection imperfections.

%All physics takes place in 1D along the X axis, where we initially start in a $20E_R$ deep lattice with $\lambda_{lat}=1064nm$.    The initial lattice loading is performed with the atoms non-interacting (i.e.\ $U=0$).  

%We then ramp up an additional lattice along the x axis with $\lambda_{lat} = 532nm$ to a depth of $20E_R$ in $200\mu s$, the relative phase of which is locked to the $\lambda_{lat}=1064nm$ lattice (see SOMs for details), creating a ``superlattice'' [cite bosons].  The phase offset between the lattices is $\approx0.85\pi$ (CHECK), creating an array of isolated double wells where the left sites are much lower in energy than the right sites (see fig. 1).  This means that all atoms initially sit on the left sites.  In this tilted configuration, where the offset and lattice depths highly supress tunelling within and between double wells, we set the interactions to the desired final value for the experiments using a magnetic field to tune the Feshbach resonance between the two spin components (see SOMs for details).

\paragraph*{Side-resolved detection sequence}

The detection sequence to determine the number of atoms on even and odd lattice sites (and hence $\mathcal{I}$) is shown in Fig.\ \ref{sequence}.  Once the evolution in the $8\, E_R$ short lattice with disorder is complete, we increase the short lattice in $10\, \mu\textrm{s}$ to $20\,E_R$ to minimize the tunneling rate.  The long lattice is then ramped up to $20\,E_R$ in $10\,\mu\textrm{s}$, while the disorder lattice is simultaneously ramped down.  With the spatial distribution of atoms now frozen in the tilted superlattice, we ramp the Feshbach field to the non-interacting ($U=0$) point in $21\,\textrm{ms}$, to ensure that there are no unwanted interaction effects during the subsequent band-mapping and imaging process.  Next, the long lattice depth is further increased to $88\,E_R$ in $20\,\mu\textrm{s}$, transferring atoms from the second band into the third band of the superlattice via an avoided crossing \cite{Trotzky12}. Any atoms on odd sites of the short lattice (the higher side of the tilted superlattice double wells) are now in the 3rd band, while atoms from even sites of the short lattice remain in the lowest band of the superlattice.  After $10\,\mu \textrm{s}$ of hold time, the short lattice is ramped to $0\,E_R$ in $200\,\mu \textrm{s}$, which is slow enough to be adiabatic with respect to inter-band transitions. Since the ordering of the bands remains unchanged during this ramp, atoms originally on even (odd) sites in the short lattice end up in the 1st (3rd) band of the long lattice.  Finally, we implement a bandmapping sequence followed by $8\,\textrm{ms}$ TOF and absorption imaging, from which we can count the number of atoms in each band directly, providing us with  $N_e$ and $N_o$.
A few atoms remaining in the second band due to an imperfect transfer are counted together with the third band.

\begin{Figure}%[h!t]
	\centering
	\includegraphics[width=84mm]{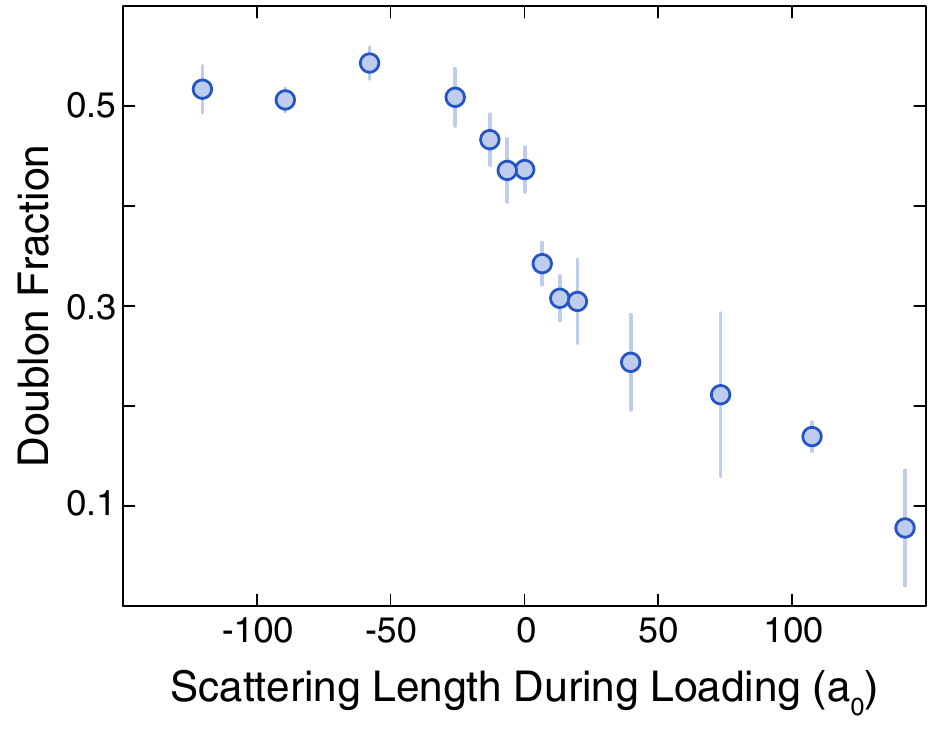} %76 %was 84mm
	\captionof{figure}{{\small{\textbf{Measured fraction of atoms on doubly occupied sites as a function of interaction strength during loading.}  The initial state prior to loading into the lattice has a temperature of $T/T_F = 0.24(2)$. Error bars show the standard deviation.}}}
	\label{Doublon_frac_U}
\end{Figure}

\paragraph*{Doublon fractions}

In order to characterize our initial state, we measure the fraction of atoms forming doubly occupied sites, termed the doublon fraction, by converting doublons into molecules by crossing the Feshbach resonance \cite{Schneider08}.  Experimentally this involves loading into a 20$E_R$ deep long lattice and $36\,E_R$ deep orthogonal lattices as per the standard sequence with loading interactions given by $a_{\text{load}}$, then ramping the long lattice to 40(1)$E_R$ in $200\,\mu$s to supress hopping in all three dimensions, which preserves the atomic distribution for the duration of the doublon detection sequence.   The Feshbach field is then ramped down to either 206.3G (= -117(2)$\textrm{a}_0$, still above the resonance) or 198.3G on the molecular side of the resonance.  When the Feshbach field ramp crosses the resonance, any pairs of atoms on the same lattice site will be converted into Feshbach molecules \cite{Koehler06}. Since the molecular transition is out of resonance of the imaging light, we can determine the doublon fraction directly by comparing the number of atoms detected at the two different Feshbach ramp endpoints.  The number of atoms is measured using standard absorbtion imaging after 6ms TOF.  Fig.\ \ref{Doublon_frac_U} shows the doublon fraction plotted against the loading interaction $a_{\text{load}}$ during the lattice ramp up.  Error bars show the standard deviation.  We quote the loading interactions in units of $\textrm{a}_0$ rather than $U/J$, as the lattice configuration at the end of the loading differs from the one during the evolution.  In the lattice configuration used for time evolution of the CDW, a scattering length of $a=7.105\,\textrm{a}_0$ corresponds to $U/J\simeq1$ .

  %For the data used to produce Fig. 3, we loaded with small repulsive interactions ($6.7(7)a_0$), which has a corresponding doublon fraction of 34(2)\%.  The doublon fraction for the three values $a_{load} = -89(2)a_0, 0(1)a_0$ and $142(1)a_0$ used for the data shown in Fig.\ 4 are 51(1)\%, 43(2)\% and 8(6)\%, respectively.

We also investigate the effect of the initial temperature   on the doublon fraction.  The initial temperature is changed by reducing the duration of the final sympathetic cooling stage and increasing the final dipole evaporation depth. This has the effect of making the cooling less efficient while keeping the final $^{40}\textrm{K}$ atom number relatively constant.  Fig.\ \ref{Doublon_frac_TTf} shows the doublon fraction plotted as a function of the temperature of the initial state, expressed in units of the Fermi temperature $T_F$, for three different loading interactions.  In the case of loading with attractive interactions or in the non-interacting case, we observe higher temperatures to reduce the resulting doublon fraction, since the increased entropy and average kinetic energy per particle reduce the average density and render the formation of doublons less favourable.  For atoms which are loaded repulsively, there are very few doublons for any temperature and thus no real dependence is observable within our error bars.

\begin{Figure}%[h!t]
	\centering
	\includegraphics[width=84mm]{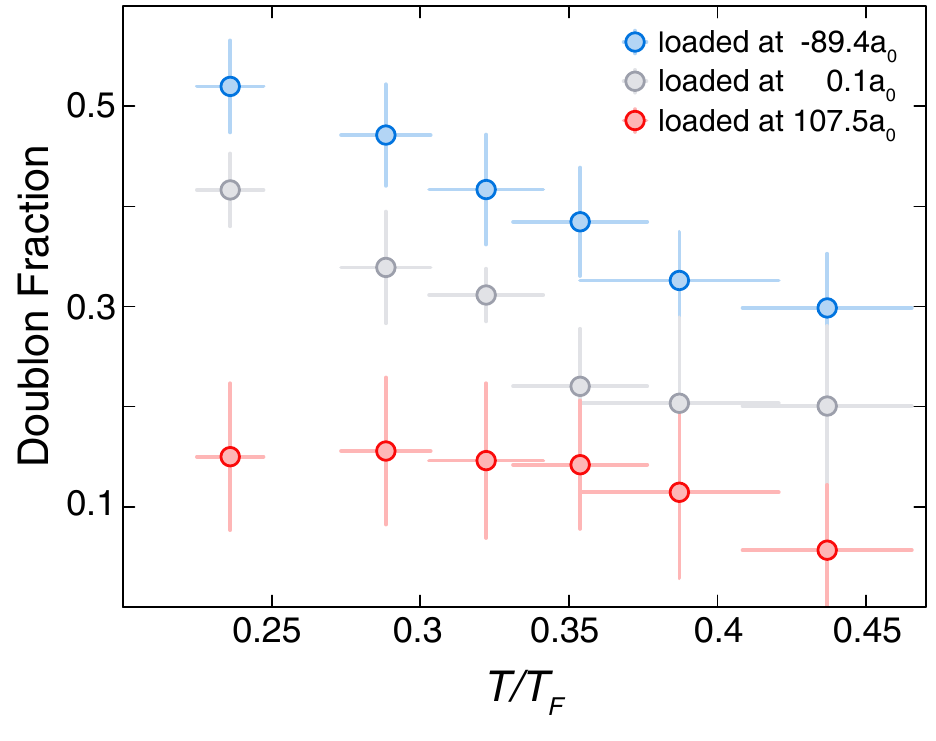} %76 %was 84mm
	\captionof{figure}{{\small{\textbf{Measured fraction of atoms on doubly occupied sites as a function of initial reduced temperature for three different loading interactions.} Error bars show the standard deviation.}}}
	\label{Doublon_frac_TTf}
\end{Figure}

\paragraph*{Imbalance vs energy density}

We investigate the effects of both loading interactions and initial temperature on the imbalance $\mathcal{I}$ in the many-body localized regime at $\Delta/J=2.3(4)$, close to the localization transition.  While changing the interactions during loading predominantly affects the number of doublons for a given temperature \cite{Hackermueller10}, changing the temperature of the initial state affects both doublon fraction and average density. Fig.\ \ref{Imb_vs_load_U} shows the effect of loading interactions on the imbalance $\mathcal{I}$ for three different interactions during evolution.

For vanishing or weak interactions during evolution, we observe no influence of loading interactions, highlighting the robustness of many-body localization in a lattice with respect to energy density.  However, for strong repulsive interactions during evolution, we see a significant effect of the loading interactions on the imbalance $\mathcal{I}$. Increasing the number of doublons by loading attractively increases the imbalance, since the additional energy scale $\Delta/J_D$ of the doublons brings the system much deeper into the localized phase. In contrast, for repulsive loading, doublons are strongly suppressed. In the limit of strong interactions (during evolution) and the absence of doublons, interacting fermions can be exactly mapped to spinless (non-interacting) fermions (see below). This is precisely what we observe in our experiment for strong repulsive loading interactions and thus vanishing doublon fraction: the imbalance observed for strong interactions during evolution converges towards the case of a non-interacting evolution.   

\begin{Figure}%[h!t]
	\centering
	\includegraphics[width=84mm]{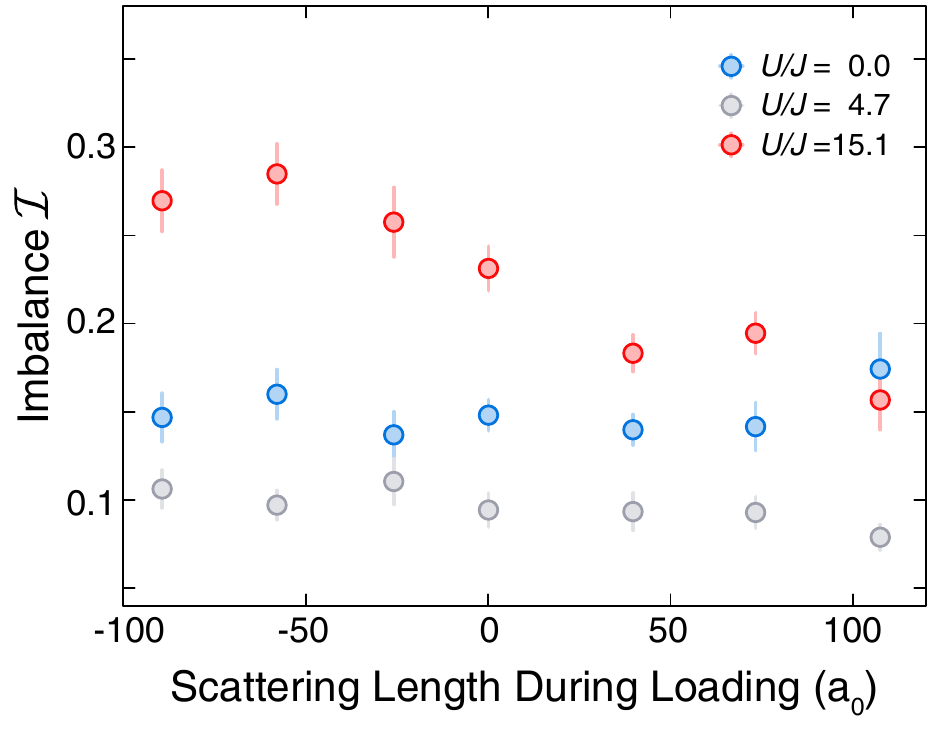} %76 %was 84mm
	\captionof{figure}{{\small{\textbf{Measured steady state imbalance as a function of loading interactions.}  Data taken with $\Delta/J=2.3(4)$ at a temperature of $T/T_F = 0.24(2)$ prior to loading.  Each $\mathcal{I}$ value is the average of the same 12 parameters as in Fig.\ 3.  Error bars show the standard deviation of the mean.}}}
	\label{Imb_vs_load_U}
\end{Figure}

\begin{figure*}[htb]
	\centering
	\includegraphics[width=180mm]{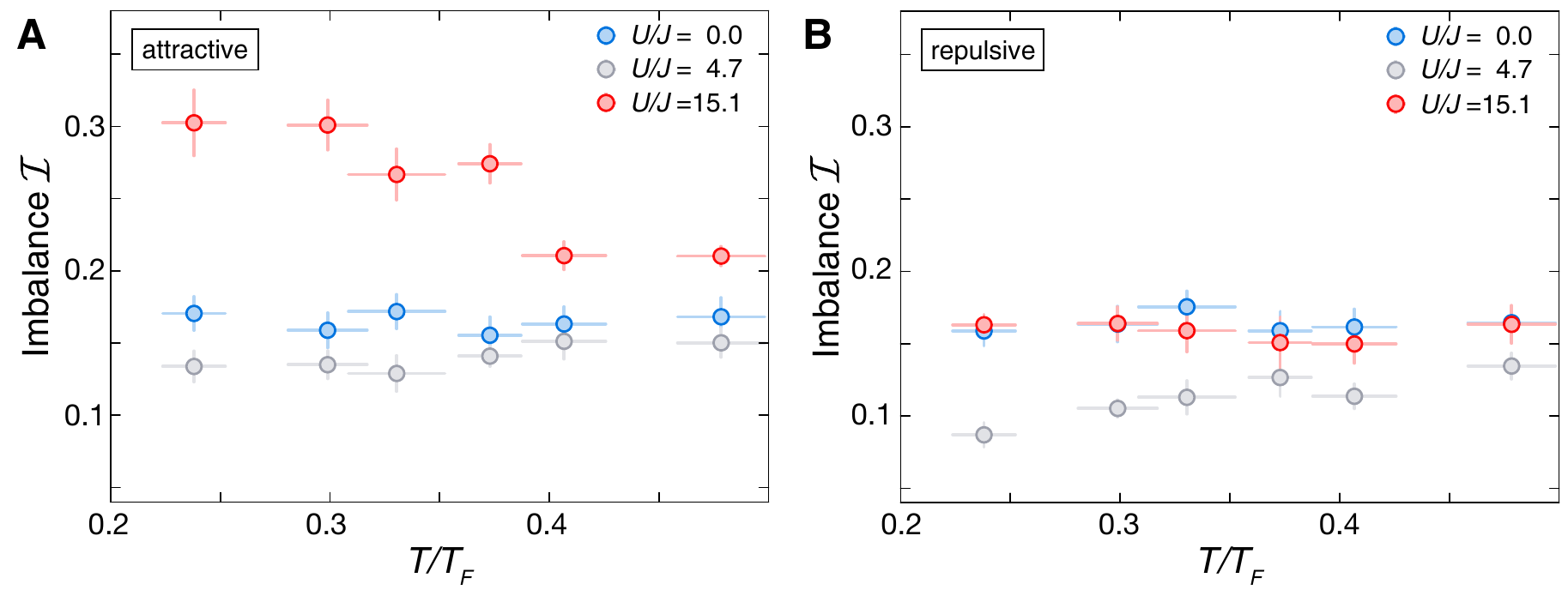} %76 %was 84mm
	\caption{{\small{\textbf{Measured imbalance as a function of initial temperature.} Data shown for \textbf{A} attractive loading and \textbf{B} repulsive loading. Each $\mathcal{I}$ value is the average of the same 12 parameters as in Fig.\ 3.  X error bars show the standard deviation, while Y error bars show the standard deviation of the mean.}}}
	\label{Imb_vs_Temp}
\end{figure*}

For intermediate interactions $U/J \approx 5$, we cannot detect any notable variation of the imbalance $\mathcal{I}$ within errors for all loading interactions. 
Additionally, in Fig.\ \ref{Imb_vs_Temp} we investigate the effect of loading temperatures on $\mathcal{I}$.  Fig.\ \ref{Imb_vs_Temp}A shows the case of attractive loading at $a_{\text{load}} = -89(2)\,\textrm{a}_0$, while Fig.\ \ref{Imb_vs_Temp}B shows the case of repulsive loading interactions $a_{\text{load}} = 101(1)\,\textrm{a}_0$.  In all cases, increased temperatures reduce the effect of interactions due to the decreasing average density, connecting the interacting cases asymptotically to the behaviour of non-interacting fermions. In the case of attractive loading, higher temperatures in addition reduce the number of doublons, which for  strong repulsive interactions during the evolution reduces the localization. 
For intermediate interactions, the effect of initial temperature is less pronounced, but for both loading interactions increasing initial  temperatures connect the minimum of localization (Fig.\ 6) to the case of no interactions , see Fig.\ \ref{Imb_vs_Temp}.

\paragraph*{Fitting of the  imbalance vs interaction curves}

We extract the minima of the imbalance as a function of interactions $U/J$ by employing the following phenomenological fitting function:

\begin{equation}
	\mathcal{I} \left(U\right) =\left|c_{1} \right| e^{-\left(\frac{U^2}{w_{1}^2}\right)}-\left|c_{2} \right| e^{-\left(\frac{U^2}{w_{2}^2}\right)}+\textrm{offset}.
\label{fitting function}	
\end{equation}
Here $c_1$, $c_2$, $w_1$, $w_2$ and $\textrm{offset}$ are free fitting parameters. 
We have implemented the  $ U \leftrightarrow  -U$ symmetry to the function by fixing the center of both Gaussians to zero. The positive amplitude Gaussian with a smaller width captures the reduction of the imbalance up to the minimum, while the decreasing contribution of the negative amplitude Gaussian resembles the increase for strong interactions. The minima of this function are used to generate the green dot-dashed lines in Fig. 4A.

\paragraph{Time evolution of the CDW in a non-interacting model}

We consider an initial state with density wave order, i.e. 
\begin{equation}
\mathcal I_0={\frac{1}{L}} \sum_l (-1)^{l}{\langle{\hat{n}_l}\rangle}_0 = {\langle{N_e-N_o}\rangle}_0/(N_e+N_o)\ne 0, 
\end{equation}
where  $L$ is the total number of sites, $N_e$ ($N_o$) is the number of fermions on even (odd) sites, and ${\langle{\ldots}\rangle}_0$ represents a trace over the initial density matrix.
This state is evolved with a (spinless) non-interacting fermionic Hamiltonian
\begin{equation}
\hat{H}=-J\sum_i (\hat{c}^\dagger_i \hat{c}^{{\phantom{\dagger}}}_{i+1} +\textrm{h.c.}) +\sum_i V_i \hat{n}_i.
\end{equation}
The disorder $V_i$ could represent true randomness or a quasi-periodic potential as in the experiment. To obtain the density wave order at later times we need to compute the expectation values of the local densities at later times, i.e.\ ${\langle{e^{i H t}\hat{n}_l e^{-i Ht}}\rangle}_0={\langle\hat{n}_l(t)\rangle}_0$.   We can write the density operators $\hat{n}_l(t)$ explicitly as 
\begin{equation}
\hat{n}_l(t)=\hat{c}_l^\dagger(t)\hat{c}^{{\phantom{\dagger}}}_l(t)=\sum_{jj'}G_{0}^*(j,l;t) G_{0}(j',l;t) \hat{c}^\dagger_j \hat{c}^{{\phantom{\dagger}}}_{j'},
\end{equation}
where $G_{0}(j,l;t)={\,\langle\,j\,|\,}e^{-i{\mathcal{H} t}}{\,|\,{l}\,\rangle\,}$ is the non-interacting propagator from point $j$ to point $l$ and ${\mathcal H}$ is the first-quantized Hamiltonian. Taking the trace over the density matrix leaves only the terms with $j=j'$ yielding
\begin{equation}
{\langle{\hat{n}_l(t)\rangle}}_0=n_0 \sum_m \left|G_{0}(2m,l;t)\right|^2,
\end{equation}
where we have assumed an initial state with particles only on even sites, i.e.\ ${\langle \hat{n}_{2m+1} \rangle}_0=0$.

Therefore the CDW or (imbalance) at time $t$ is
\begin{equation}
{\mathcal I}(t)={1\over L}\sum_{l,m}(-1)^l \left|G_0(2m,l;t)\right|^2,
\label{eq:imbT}
\end{equation}
which can be calculated numerically

\paragraph{Details of system modelling via exact diagonalisation}

In order to calculate the red line in Fig.\ 3, we employ Eq.~(\ref{eq:imbT}) to calculate the time evolution of the initial CDW  in a single tube in the presence of the harmonic trap. The initial CDW is modeled as a collection of randomly distributed localized particles, whose distribution follows a Gaussian with $(\sigma_x=21\,\mu\textrm{m})$ and matches the initial imbalance of $\mathcal{I}_0=0.96(2)$ observed in the experiment. We extract the stationary imbalance by averaging the imbalance in the evolution time interval from $15\, \tau$ to $30\,\tau$. In addition, the calculations are averaged over twelve disorder realizations corresponding to twelve equally spaced phases  $\phi\in \left[0,2\pi\right]$ in Eq.\ (\ref{AA_hamiltonian}).

In order to correctly model the dynamic evolution shown in Fig.\ \ref{time_trace_non_int}, we additionally integrate the results over many individual 1D tubes with tube-dependent hoppings in order to fully model the experimental situation. The widths and amplitudes of the corresponding Gaussian distributions were chosen in accordance with the experimental cloud sizes of $\sigma_{x,y}=21\,\mu\textrm{m}$ and $\sigma_z=5.5\,\mu\textrm{m}$ that were obtained via in-situ imaging. We included both the hopping perturbation resulting from the disorder potential $V_i = \Delta \cos(2\pi \beta i + \phi)$ as well as the effect of the finite waist of the lattice beams. The strength and phase difference $\phi_{diff}$ of the disorder induced hopping modulation are extracted from fits to the exact position dependent hopping rates.

\paragraph{Stationary imbalance value in the localized phase}

We can give an analytic estimate for the steady state value at long times in the localized regime. Time averaging the wave function of a particle originating from a single site should lead to a form which is  exponentially decaying over a localization length $\xi$. Hence, we can approximate
$
\left|G_0(l',l;t)\right|^2\sim {1\over \xi} \left(1+e^{i\pi l'}\right)e^{-|l-l'|/\xi}$, 
where the prefactor $1+e^{i\pi l'}$ ensures that the initial site of the wave function is an even site. Note that this form holds for $\xi\gg d$, i.e.\ a localization length much larger than a lattice constant $d=\lambda_s/2$. 
Inserting this expression into Eq. (\ref{eq:imbT}) for the density wave order yields

\begin{Figure}%[h!t]
	\centering
	\includegraphics[width=84mm]{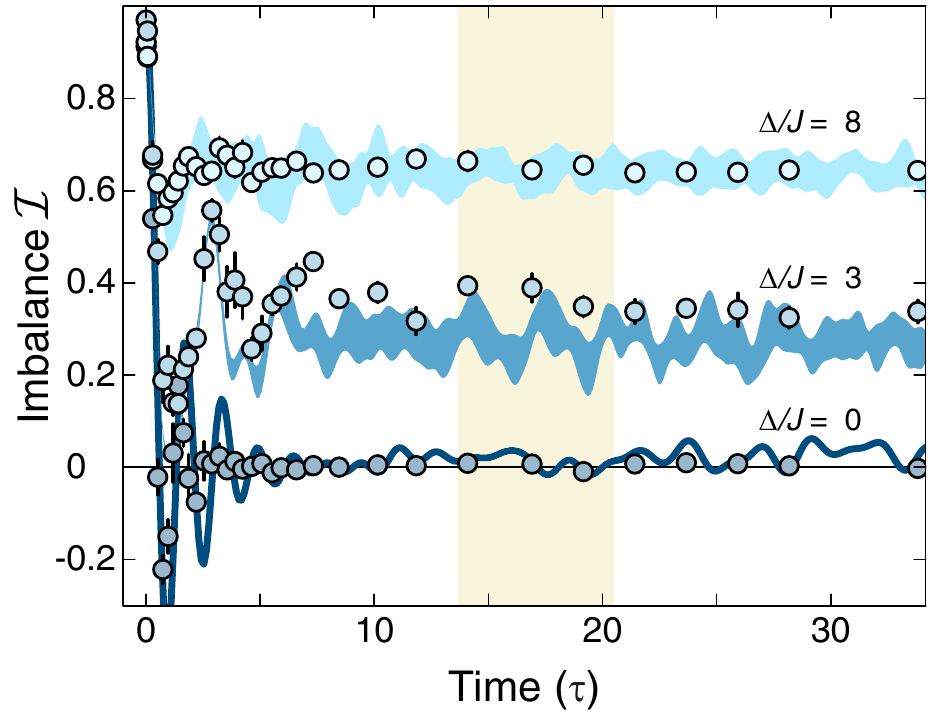} %76 %was 84mm
	\captionof{figure}{{\small{\textbf{Time evolution of a non-interacting initial charge-density wave.} We experimentally measure (circles) the time evolution for various disorder strengths and compare it to an exact diagonalisation simulation, taking into account the trapping potential and the variations in tunneling. Experimental error bars show the standard deviation of the mean, while the shaded region indicates the standard deviation of simulation runs for various disorder phases. }}}
	\label{time_trace_non_int}
\end{Figure}

\begin{equation}
\begin{split}
{\mathcal I}(\infty)\sim{1\over L}\sum_{ll'}\left(e^{-i\pi(l-l')} +e^{-i\pi l}\right){1 \over \xi}e^{-|l-l'|/\xi} \\=
{1\over L}\sum_{r,{\bar l}}\left(e^{-i\pi r} +e^{-i{\pi\over 2}r}e^{-i\pi {\bar l} }\right){1 \over \xi}e^{-|r|/\xi},
\end{split}
\end{equation}
where in the last equality we changed to a sum over relative $r$ and center of mass coordinates ${\bar l}$. The second term vanishes and we find (for large $\xi\gg d$)
\begin{equation}
{\mathcal I}(\infty)\sim{1\over L}\sum_r e^{-i\pi r} {1 \over \xi}e^{-|r|/\xi}\sim\frac{1}{\xi^2},
\end{equation}
This formula establishes the connection between the CDW stationary value and the localization length close to the transition in a non-interacting system.

\paragraph{Mapping from hard-core to non-interacting fermions}

We consider the dynamics of the (spin-$1/2$) Hubbard model in the limit of infinite (on-site) $U$ when the initial state has no doubly-occupied sites. In this case, the time evolution can be mapped to that of non-interacting spinless fermions.    

In the limit $U/J\to\infty$ the Hilbert sub-space with no doubly-occupied sites is closed under unitary time evolution. Double occupancies cannot be generated in the dynamics while conserving energy because they incur an infinite energy cost. Hence, we can replace the interaction with a hard-core constraint implemented by the operator $\hat{P}_{G}$, which projects out the doubly-occupied subspace,
\begin{equation}
\hat{H}=-J\sum_{i{\sigma}}\left[\hat{P}_{G}\hat{c}^\dagger_{{\sigma}, i}\hat{c}^{{\phantom{\dagger}}}_{{\sigma}, i+1} \hat{P}_{G} +\textrm{h.c.}\right]+ \sum_{i} V_i \hat{n}_{i}.
\label{Hhc}
\end{equation}
Let us denote the ordering of the fermion spins from left to right on the chain with $\{\sigma\}\equiv\{\sigma_1,\sigma_2,\ldots,\sigma_N\}$, where $N$ is the total fermion number. $\{{\sigma}\}$ is an invariant of the dynamics because in order to exchange the spins of two particles with opposite spin orientation, the particles need to meet on the same site, which is prohibited by the hard-core constraint.  Now, suppose we start the time evolution with an initial state having a definite spin ordering $\{{\sigma}\}$. Since this is an invariant of the dynamics, we can describe the time evolution within this sector with a Hamiltonian $\hat{H}[\{{\sigma}\}]$ in which the spin ordering enters as a parameter. The Hilbert space in which this Hamiltonian acts can be faithfully represented by the Fock space of spinless fermions ${\,|\,{n_1,\ldots,n_L}\,\rangle\,}_{\{{\sigma}\}}$, because given the fermion positions their spins are uniquely determined by the configuration $\{{\sigma}\}$ which parameterizes the space. Hence we can write the Hamiltonian which acts within the $\{{\sigma}\}$ subspace as a Hamiltonian of spinless fermions
\begin{equation}
\hat{H}[\{{\sigma}\}]=-J\sum_{ij} \left[\hat{f}^\dagger_i \hat{f}^{{\phantom{\dagger}}}_{i+1}+\textrm{h.c.}\right]+\sum_iV_i\hat{f}^\dagger_i \hat{f}^{{\phantom{\dagger}}}_i.
\label{Hf}
\end{equation}
Note that the projection of doubly-occupied sites is automatically taken into account using the representation with spinless fermions. 

The mapping of the time evolution within a given spin-ordering sector $\{{\sigma}\}$ can also be understood as a result of a  non-local unitary change of basis 
\begin{eqnarray}
U[\{{\sigma}\}]=\prod_{j=1}^L \exp\left( -i \varphi_{\{{\sigma}\}}[ N_{<j}+1]\hat{S}^x_j \right),
\end{eqnarray}
where $N_{<j}$ denotes the particle number to the left of point $j$, $\hat{S}^x_j= \frac{1}{2} \hat{c}^\dagger_{{\alpha} j}{\sigma}^x_{{\alpha}{\alpha}'} \hat{c}^{{\phantom{\dagger}}}_{{\alpha}'j}$ and
\begin{equation}
\varphi_{\{{\sigma}\}}[n_j] = \begin{cases} 0, & \mbox{if } {\sigma_n=\uparrow }\\ \pi, & \mbox{if } {\sigma_n=\downarrow}\end{cases}.
\end{equation}
In other words, the transformation counts the number of particles to the left of point $j$ to find its order in the chain $n_j$. It then rotates the spin by $\pi$ around the $x$ axis if in the configuration $\{{\sigma}\}$ the $n_j$-th spin is $\downarrow$. The transformed fermions are then all $\uparrow$ and we can drop the spin index.  

The dynamics is thus described by the same Hamiltonian of non-interacting spinless fermions regardless of the initial ordering of the spins $\{{\sigma}\}$. 
In particular, the time dependence of the CDW order can be generated by unitary evolution of a spinless non-interacting fermion model and is independent of the initial spin ordering.

\paragraph{Time evolution of the CDW in an interacting model}

We consider the following ensemble of many-body states as initial states for the numerical calculation:
\begin{equation}
    |\Psi_0\rangle = \prod_m\left(\hat{c}^{\dag}_{2m\uparrow}\right)^{n_{2m\ua}}\left(\hat{c}^{\dag}_{2m\downarrow}\right)^{n_{2m\da}}|\Omega\rangle,
\end{equation}
where $|\Omega\rangle$ describes the vacuum and $n_{i\a}=0,1$. In order to generate the ensemble of initial states, the occupations $n_{i\a}=0,1$ are chosen from an appropriate joint probability distribution $p(n_\ua,n_\da)$ such that the average site occupation and the average doublon occupation are fixed.  This state is time evolved with the interacting Hamiltonian
\begin{equation}
    \hat{H}=-J\sum_{i,\sigma} (\hat{c}^{\dag}_{i,\sigma} \hat{c}^{\phantom{\dag}}_{i+1,\sigma} +\textit{h.c.}) +\sum_{i, \sigma} V_i \hat{n}_{i, \sigma} + U \sum_{i}\hat{n}_{i,\uparrow}\hat{n}_{i,\downarrow}
\end{equation}
using the time-evolving block decimation (TEBD)~\cite{vidal:2004, schollwoeck:2011} on matrix product states implemented with the ITensor Library~\cite{itensor}.

For our numerical simulations we  use a quasi-periodic potential $V_i = \Delta \cos(2\pi \beta i + \phi)$ with $\beta\approx 0.721$ as in the experiment. However, we do not take into account the periodic trap, but rather consider a system with $L=40$ sites and open boundary conditions. When simulating the time evolution for a given set of system parameters, i.e.\ $\Delta$, $U$, site occupation, and doublon density, we average over both the initial state and the disorder realization (through sampling of the random phase $\phi$). The total number of runs for a given set of parameters is $\approx 100=10\times 10$, i.e.\ about 10 initial states for every disorder realization.
For the singular value decomposition, we use a cutoff of $10^{-5}$ and allow for a maximum bond dimension of $m_{\rm max} = 100$. Using a time step of $\Delta t = 0.1 \;[\hbar / J]$, this leads to a maximum truncation error $\lesssim 10^{-4}$.

\paragraph{Fluctuations of the imbalance}

In the main text we presented TEBD (i.e.\ DMRG) calculations showing logarithmic growth of the entanglement entropy with time, which is characteristic of the many-body localized state.
The entanglement entropy, however, is a non-local quantity and therefore not directly measurable. In this section we show that the same information about the many-body localized state can be inferred from the temporal fluctuations of the measured imbalance $\mathcal{I}$ (CDW order parameter) around its stationary value.
 
We focus on the system deep in the localized phase, where it can be described using an effective model of local integrals of motion. For simplicity we initialize the system in a CDW state with exactly one particle on every even site. We now partition the lattice into two-site dimers and, as a starting point, take the dimers to be disconnected.  We define
 $\sigma^z_i=\hat{n}_{2i}- \hat{n}_{2i+1}$ and $\s^x_i=\sum_\s \hat{c}\yd_{2i}\hat{c}^{\phantom{\dag}}_{2i+1}+\textrm{h.c.}$, which within the accessible Hilbert space act like Pauli spin operators. In terms of these operators the imbalance (or CDW order) reads
\begin{equation}
    \mathcal{I} = \frac{1}{L}\sum_i \langle\sigma_i^z\rangle.
\end{equation}

For a system with strongly localized particles (i.e.\ $\xi_{*}\ll 1$) we will continue to use an effective model of decoupled dimers, where from now on $L$ denotes the number of dimers. We take into account the tail of the single particle wave-function, which  extends outside of the dimers, only through the interaction it induces between the local spin operators. Hence an effective model is  
\begin{equation}
    H = \sum_i \vec{h}_i\cdot\vec{\sigma}_i + \sum_{i,j}V_{ij}\tilde{\sigma}_i^z\tilde{\sigma}_j^z,
\end{equation}
where $\vec{h}_i = J \hat{x} + \epsilon_i\hat{z}$, $\tilde{\sigma}_i^z = (\vec{h}_i\cdot\vec{\sigma}_i)/|\vec{h}_i|$ are the integrals of motion and the interaction is given by $V_{ij} \sim U \exp(-|x_i - x_j|/\xi_{*})$.

For the non-interacting case, only the first term of the effective model is non-zero, and we directly find the expectation value of the spin operator on site $i$
\begin{equation}
    \langle \sigma_i^z(t)\rangle = \cos^2 \theta_i + \sin^2\theta_i \cos(\omega_i t),
\end{equation}
where $\omega_i = \sqrt{\epsilon_i^2 + J^2}$ and $\cos^2\theta = \epsilon_i^2 / \omega_i^2 \approx 1 - (J/\epsilon_i)^2$. Thus, we find the time-averaged stationary value of the imbalance is $\mathcal{I}={1\over L}\sum_i \cos^2\t_i$. The temporal fluctuations around this average are
\begin{equation}
\begin{split}
  \av{\delta\mathcal{I}(t)^2}_T &=\left\langle\frac{1}{L^2}\sum_{i,j} \sin^2\theta_i\sin^2\theta_j\cos(\omega_i t)\cos(\omega_j t)\right\rangle_T \\ &\approx \left\langle\frac{1}{L^2}\sum_i \sin^4\theta_i \cos^2 (\w_i t) \right\rangle_T \sim \frac1L,
\end{split}
\end{equation}
where $\av{\ldots}_T$ represents averaging over a long time window $T$.
We conclude that without interactions the fluctuations of the imbalance remain on average constant in time   with $\delta\mathcal{I}_{\rm rms} \equiv \sqrt{\av{\delta\mathcal{I}(t)^2}_T} \sim 1/\sqrt{L}$.

When interactions are present the local contribution to the density wave $\av{\s^z_i(t)}$ is affected by the time evolution of the other spins. The pseudo spin $\s^z_i$ gradually becomes entangled with a growing number of other spins. Specifically, because the interactions $V_{ij}$ decay exponentially with distance, the number of spins that become significantly entangled with a given spin at time $t$ is $l(t) \sim S(t) = S_{\rm offset} + s_* \ln (Ut/\hbar)$. We will see below how this affects the fluctuations of $\av{\s^z_i}$ and ultimately contributes to the temporal fluctuations of the order parameter $\mathcal{I}$.

Let us calculate the reduced density matrix of the pseudo spin at site $i$.
The system starts the time evolution in a product state
\begin{equation}
    |\Psi_0\rangle = \sum_{\sigma_1, \ldots, \sigma_l}\prod_{j=1}^{l(t)}A^j_{\sigma_i}|\{\sigma\}\rangle 
\end{equation}
with $A_{\uparrow}^j = \cos (\theta_j/2)$ and $A_{\downarrow}^j = \sin (\theta_j/2)$. We can formally write  the time-dependent density matrix as
\begin{equation}
    \tilde{\rho}(t) = \sum_{\{\sigma\}}\sum_{\{\sigma'\}}\prod_j^{l(t)} A_{\sigma_j}^j (A_{\sigma'_j}^j)^* e^{-i(E[\{\sigma\}] - E[\{\sigma'\}]t)}  |\{\sigma\}\rangle\langle\{\sigma'\}|.
\end{equation}
Here, $E[\{\sigma\}] = \sum_i h_i\sigma_i + \sum_{i,j}V_{ij}\sigma_i\sigma_j$. We can now trace out all but one site to obtain the reduced density matrix of a single pseudo-spin at site $i$ in the basis of the eigenvalues of $\tilde{\s}^z_i$:
\begin{eqnarray}
    \tilde{\rho}_{\uparrow\uparrow} &=& \cos^2\theta_i/2\\
    \tilde{\rho}_{\downarrow\downarrow} &=& \sin^2 \theta_i/2\\
    \tilde{\rho}_{\uparrow\downarrow} &=& \tilde{\rho}_{\downarrow\uparrow}^* = \sin\theta_i \, \left[{\frac{1}{N_f(t)}}\sum_{n=1}^{N_f(t)}e^{-i \omega_n t} \right],
\end{eqnarray}
where $\omega_n = E[\uparrow, \{\sigma\}] - E[\downarrow,\{ \sigma'\} ]$. $\{\s\}$ and $\{\s'\}$ represent  states of all other sites except site $i$. The number of frequencies involved $N_f(t)$ is roughly $2^{l(t)}$, i.e.\ all possible interactions between the $l(t)$ spins that are significantly entangled with the observed spin at time $t$. More precisely, the number of frequencies $N_f(t)=e^{S(t)}\approx e^{s_*\ln (Ut/\hbar)}$ measures the size of the entangled Hilbert space at the observation time.  %The important consequence for our purpose  is that noise produced by a large number $N_f$ independent frequencies will be suppressed by $1/\sqrt{N_f}$.

To find the time dependence of the CDW we have to transform back from the basis of $\tilde{\s}^z_i$ eigenvalues to the basis of $\s^z_i$.  This is achieved with a rotation around the ${\tilde{\s}}^y$ axis by $\theta_i$
\begin{equation}
    \langle \sigma^z(t)\rangle = {\rm Tr}(\rho(t)\sigma^z) = \cos^2\theta_i + \sin^2\t_i \frac1{N_f(t)}\sum_{n=1}^{N_f(t)}\cos(\omega_n t).
\end{equation}
Hence, the fluctuations of the local imbalance between an even site and neighboring odd site behaves as
\begin{equation}
    \delta\sigma^z_{\rm rms}(t) \sim e^{- \half S(t)} \sim \Big(\frac{\hbar}{Ut}\Big)^{\half s_*},
\end{equation}
while the fluctuation of the global imbalance (CDW order) is further suppressed by a factor of $1/\sqrt{L}$
\begin{equation}
    \delta\mathcal{I}_{\rm rms}(t) \sim \frac{1}{\sqrt{L}}\Big(\frac{\hbar}{Ut}\Big)^{\half{s_*}}.
\end{equation}
We see that the decay of the fluctuations is intimately connected to the $\ln t/\tau$ growth of the entanglement entropy, as it also `measures' the number of spins coupled through the interactions after time $t$.

\begin{Figure}%[h!t]
	\centering
	\includegraphics[width=84mm]{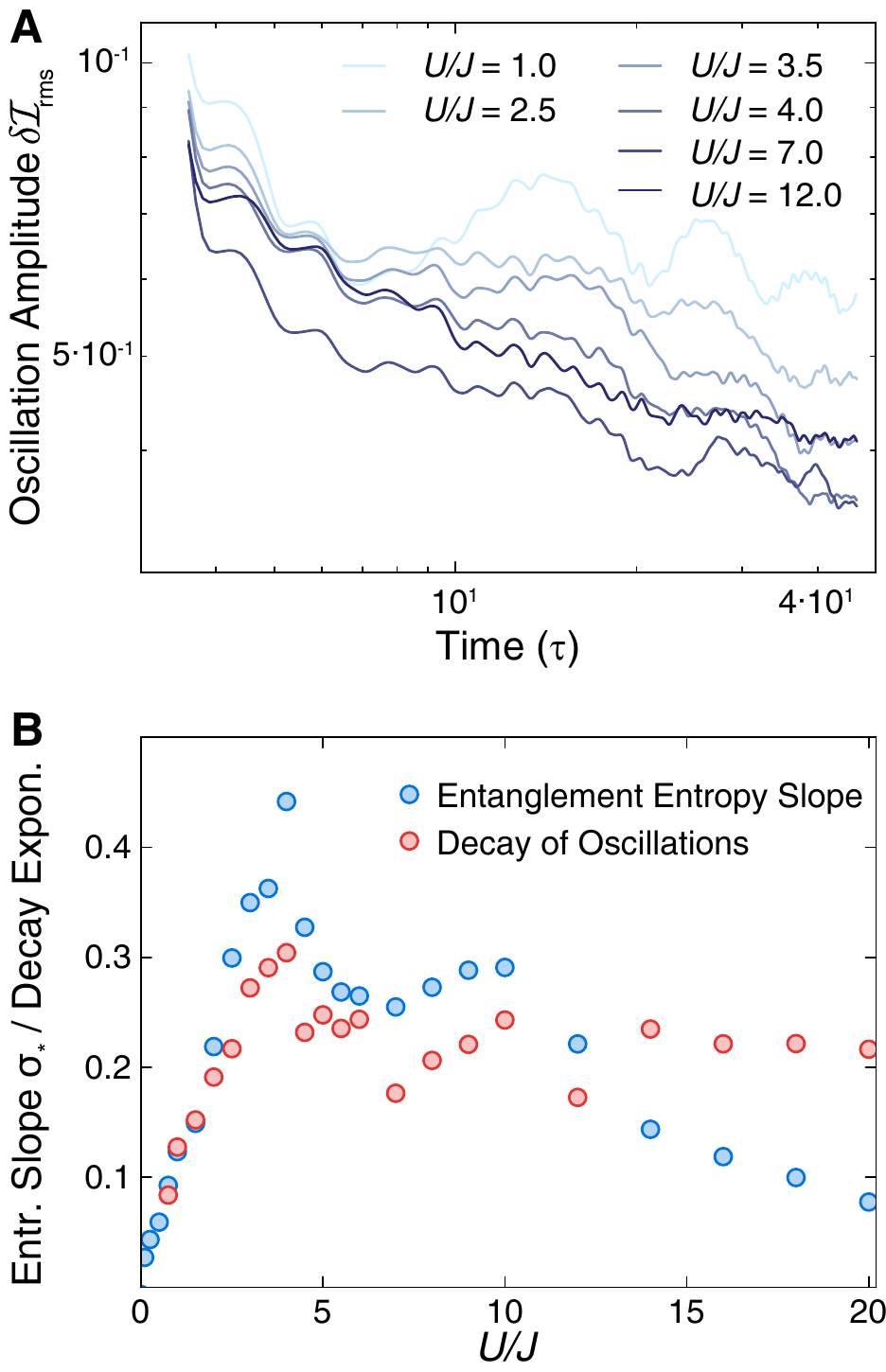} %76 %was 84mm
	\captionof{figure}{{\small{\textbf{Decay of imbalance oscillations and entanglement entropy growth.}  \textbf{A} shows the time evolution of the imbalance with decreasing oscillation amplitudes for different interaction strengths for a system with $30\%$ doublons, exhibiting a power law decay. \textbf{B} shows the connection of the decay exponents of the imbalance oscillation amplitudes and the slope of the logarithmic growth of entanglement entropy as a function of interaction strength.}}}
	\label{ent_osc_theory}
\end{Figure}

The above analysis is of a simplified effective model. However, the main conclusions are supported by direct simulation of the Hubbard model on the quasi-periodic lattice using time dependent DMRG. Fig. \ref{ent_osc_theory}A shows the temporal noise of the imbalance as a function of the time for different values of the interaction strength $U/J$. The fluctuations are measured by averaging them over a time window of $T = 7\,\tau$ around the time $t$. The results fit well to a power law decay. Fig. \ref{ent_osc_theory}B compares the fitted exponent $s_*$ to the slope of the logarithmic growth of the entanglement entropy showing the direct correspondence between the two.

Thus, we conclude that measurements of the temporal fluctuations of the CDW order provide a viable experimental route to determine the bare localization length $\xi_*$ and distinguish the many-body localized state from an Anderson localized state of non-interacting particles. 
In the present experiment, this is, however, not possible, as the unavoidable averaging over many tubes suppresses the fluctuations below the detection limit after only few oscillations. For future experiments, a single tube, or even single-site resolution would be desirable to overcome this limitation. We also note that the temporal fluctuations of the expectation value are different from shot-to-shot fluctuations at a given time which reflect the quantum uncertainty of the observable and would be finite even in the infinite-time limit.

\end{multicols}

\end{document}